\documentclass{aastex}
\usepackage{natbib,epsfig,emulateapj5,rotate}

\tightenlines
\begin{document}

\newcommand{\be}{\begin{eqnarray}}
\newcommand{\ee}{\end{eqnarray}}
\newcommand{\vperp}{V_{\perp}}
\newcommand{\pdfi}{f_{\rm S}}
\newcommand{\cdfi}{F_{\rm S}}
\newcommand{\im}{S_{\rm max}}
\newcommand{\iave}{S_{\rm avg}}
\newcommand{\iavep}{S_{\rm avg}^{\prime}}
\newcommand{\ssys}{S_{\rm sys}}
\newcommand{\snr}{S/N}
\newcommand{\npol}{N_{\rm pol}}
\newcommand{\Smax}{S_{\rm max}}
\newcommand{\Save}{\langle S \rangle}
\newcommand{\Dmax}{D_{\rm max}}
\newcommand{\DM}{{\rm DM}}
\newcommand{\DtDM}{\Delta t_{\DM}}
\newcommand{\DtdDM}{\Delta t_{\delta \DM}}
\newcommand{\DtDnu}{\Delta t_{\rm \Delta\nu}}
\newcommand{\DtISS}{\tau_d}
\newcommand{\Dnu}{\Delta\nu}
\newcommand{\dDM}{\delta\DM}
\newcommand{\Nh}{N_{\rm h}}
\newcommand{\Dnuiss}{\Delta\nu_{\rm d}}
\newcommand{\Dtiss}{\Delta t_{\rm d}}
\newcommand{\SM}{{\rm SM}}
\newcommand{\thetad}{\theta_{d}}
\newcommand{\thetaiso}{\theta_{\rm iso}}
\newcommand{\ld}{\ell_{d}}
\newcommand{\thetaisor}{\theta_{\rm iso}}
\newcommand{\lr}{\ell_{r}}
\newcommand{\dkpc}{D_{\rm kpc}}
\newcommand{\viss}{V_{\rm ISS}}
\newcommand{\vissk}{V_{\rm ISS,5/3,u}}
\newcommand{\vsperp}{{V_{\rm s}}_{\perp}}
\newcommand{\Niss}{N_{\rm ISS}}
\newcommand{\tsp}{S/N_{\rm min}}

\title{Searches for Giant Pulses from Extragalactic Pulsars}
\author{M.\ A.\ McLaughlin\altaffilmark{1} \& J.\ M.\ Cordes\altaffilmark{2}}
\altaffiltext{1}{Jodrell Bank Observatory, University of Manchester, Macclesfield, Cheshire, SK11 9DL, UK}
\altaffiltext{2}{Astronomy Department, Cornell University, Ithaca, NY 14853}

\begin{abstract}

We discuss the giant-pulse phenomenon exhibited by
pulsars and the distances to which giant pulses might be detected from
extragalactic pulsars. We describe the conditions under which a
single-pulse search is more sensitive than a standard periodicity search.
We find that, for certain pulse-amplitude distribution power laws and time
series lengths, single-pulse searches can be superior.  We present the
results of searches toward several extragalactic targets, including M33,
the LMC (PSR~B0540$-$69) and several other galaxies. While we have not
conclusively detected giant pulses from any of these targets, these
searches illustrate the methodology of, issues related to and difficulties
in these types of searches.

\end{abstract}
\keywords{pulsars, radio pulses, M33, PSR~B0540$-$69}

\section{Introduction}

While pulsar flux densities are typically quoted as averaged quantities,
individual pulses in fact follow a distribution of pulse intensities, with
some pulses having intensities of more than 10 times the mean
\cite{hesse74}. Many pulsars show approximately normal distributions of
pulse intensities about a mean, while some show bimodal distributions,
with a finite probability of zero power (e.g.~Lyne \& Graham-Smith 1998).
Other pulsars show asymmetric distributions, with a long tail towards high
values of pulse intensity. The Crab pulsar (B0531+21) is an extreme
example of this case, with a `giant' pulse more than 10 times the mean
intensity occurring once every 1000 pulses \cite{hankins75,lund95}. Such pulses
have also been detected from the the millisecond pulsars
B1821$-$24 and B1937+21 \cite{cognard96,romani01} and, very recently, from the
 young pulsar B0540$-$69 \cite{jr03}.
We explore how this giant-pulse
phenomenon can be exploited to detect pulsars that may not be detectable
through standard periodicity searches. While the main motivation for such
a search is the detection of distant, Crab-like pulsars in other galaxies,
searches for isolated, dispersed pulses may also be more sensitive than
periodicity searches to pulsars seen through the Galactic scattering disk,
pulsars with very fast spin periods and pulsars in short binary orbits.
There is also the possibility of detecting other classes of radio bursting
objects, as discussed in Cordes \& McLaughlin (2003) (hereafter Paper I).

In Paper I, we discussed issues related to the detection of fast radio
transients from astrophysical sources. We described search algorithms and
methodology and in particular discussed the impact of interstellar
scintillation and scattering on such searches. In this paper, we apply the
methodology of Paper I to searches for giant pulses from extragalactic
pulsars. We describe our searches for single radio pulses from M33,
PSR~B0540$-$69 in the Large Magellanic Cloud, and several other galaxies.
While no giant pulses have been conclusively detected in these searches,
we hope that they can serve as blueprints for future, more sensitive
searches for radio transients. Although giant pulses from B0540$-$69 have
recently been detected by Johnston \& Romani (2003; hereafter JR03), we
present the results of our earlier null search here and discuss the implications
of our search for the giant-pulse properties of B0540$-$69.

The outline of this paper is as follows. In \S\ref{sec:giant}, we describe
the giant-pulse phenomenon seen in the Crab pulsar and a few others. In
\S\ref{sec:eg}, we explore the capabilities of single-pulse searches for
detecting distant pulsars and compare the sensitivities of single-pulse
and standard periodicity searches.  In \S\ref{sec:targets}, we present
results for several targets, including the nearby spiral galaxy M33, 
B0540$-$69 and a few other galaxies. 
Conclusions and a look to
the future are offered in \S\ref{sec:conclusions}.

\section{The Giant-Pulse Phenomenon} \label{sec:giant}

In
one hour of observation, the largest measured peak pulse flux of the Crab
is roughly $\sim$~10$^{5}$~Jy at 430~MHz for a duration of roughly
100~$\mu$s \cite{hankins75}, corresponding to an implied brightness
temperature of 10$^{31}$ K. Recently, pulses with flux $\sim$ $10^{3}$~Jy
at 5~GHz for a duration of only 2~ns have been detected from the
Crab \cite{hankins03}. These `nano-giant' pulses imply brightness
temperatures of 10$^{38}$~K, by far the most luminous emission from any
astronomical object. For many years, this phenomenon was thought to be
uniquely characteristic of the Crab. However, giant pulses have since been
detected from the Crab-like pulsar B0540$-$69 and the
millisecond pulsars PSRs~B1937+21 and B1821$-$24.  For B0540$-$69, the largest
measured peak pulse flux in one hour is approximately 4~Jy \cite{jr03}. From 
both millisecond pulsars, the largest measured peak pulse flux in one hour is
$\sim$ 10$^{3}$~Jy at 430~MHz \cite{cognard96,romani01}.

As shown in Figure~\ref{fig:temp}, the giant pulses detected from the
Crab and PSRs~B0540$-$69, B1937+21 and B1821$-$24 represent the most luminous pulsed
emission detected from any pulsar. While PSRs~B1937+21 and B1821$-$24
are `recycled' millisecond pulsars with very different ages, spin periods
and surface magnetic field strengths than the Crab and PSR~B0540$-$69, all four pulsars have
similarly high values of magnetic field at the light cylinder
 $B_{lc}$  ($B_{lc} \approx 10^{8.46}
P^{-\frac{5}{2}}\dot{P}^{\frac{1}{2}}$~G, where $P$ is pulsar period is seconds and 
$\dot{P}$ is period derivative). This suggests that giant-pulse physics may
depend on conditions there, rather than at the stellar surface. All four
pulsars also show high-energy emission at the same phase as their giant
pulses, suggesting a relationship between the two phenomena and that the
giant-pulse emission process occurs high in the magnetosphere. In
Table~\ref{tab:blcs} we list the radio pulsars with the highest values of the
magnetic field at the light cylinder. 

\medskip
\epsfxsize=9truecm
\epsfbox{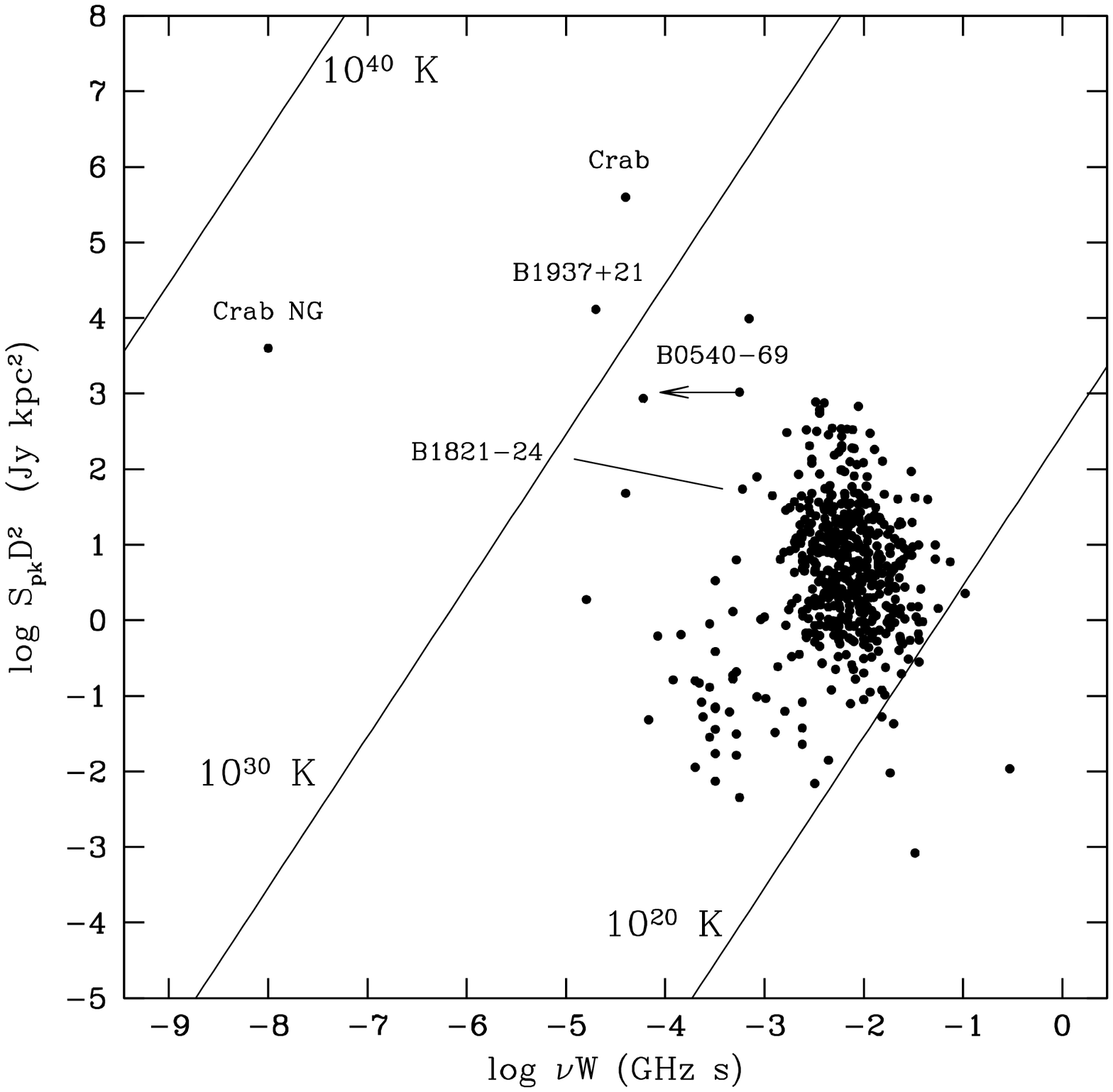}
\figcaption{
\label{fig:temp}
A log-log plot of the product of peak flux $S$ in Jy and the square of the
distance $D$ in kpc vs. the product of frequency $\nu$ in GHz and pulse
width $W$ in s for the `nano-giant' pulses detected from the Crab
\cite{hankins03}, the giant pulses detected from the Crab and PSRs~B0540$-$69, B1937+21 and
B1821$-$24 and single pulses from all pulsars with flux, distance and
pulse width listed in the Princeton Pulsar Catalog \cite{ppcat}.  Lines of
constant brightness temperature $T = S D^{2}/2k(\nu W)^{2}$ are shown,
where $k$ is Boltzmann's constant.
}
\bigskip

Lundgren et al. (1995) found that the distribution of Crab pulse fluxes is
bimodal, with separate components of the probability distribution function
(PDF) describing the normal and giant pulses. As shown in
Figure~\ref{fig:ampdist}, the flux distribution of giant pulses, defined
to be those with fluxes exceeding the detection threshold determined by
the Crab nebula, can be roughly described by a power law for flux
densities above the turnover at $\sim$~200~Jy. The differential
distribution $dN/dS \propto S^{-\alpha}$ with index $\alpha \sim$~3.5.
Similarly, the giant-pulse intensity distribution for PSR~B1937+21 can be
described by a power law of index 2.8 \cite{cognard96}. Unlike the Crab's
giant pulses, however, the giant pulses of PSR~B1937+21 appear to be the
extreme of a continuous distribution of pulse intensities. The giant
pulses of PSR~B1821$-$24 are less well-studied but seem to be adequately
described by a power law of index 3 $-$ 5 \cite{romani01}. Too few pulses
have been detected from PSR~B0540$-$69 to constrain their flux density 
distribution \cite{jr03}.

\medskip
\epsfxsize=9truecm
\epsfbox{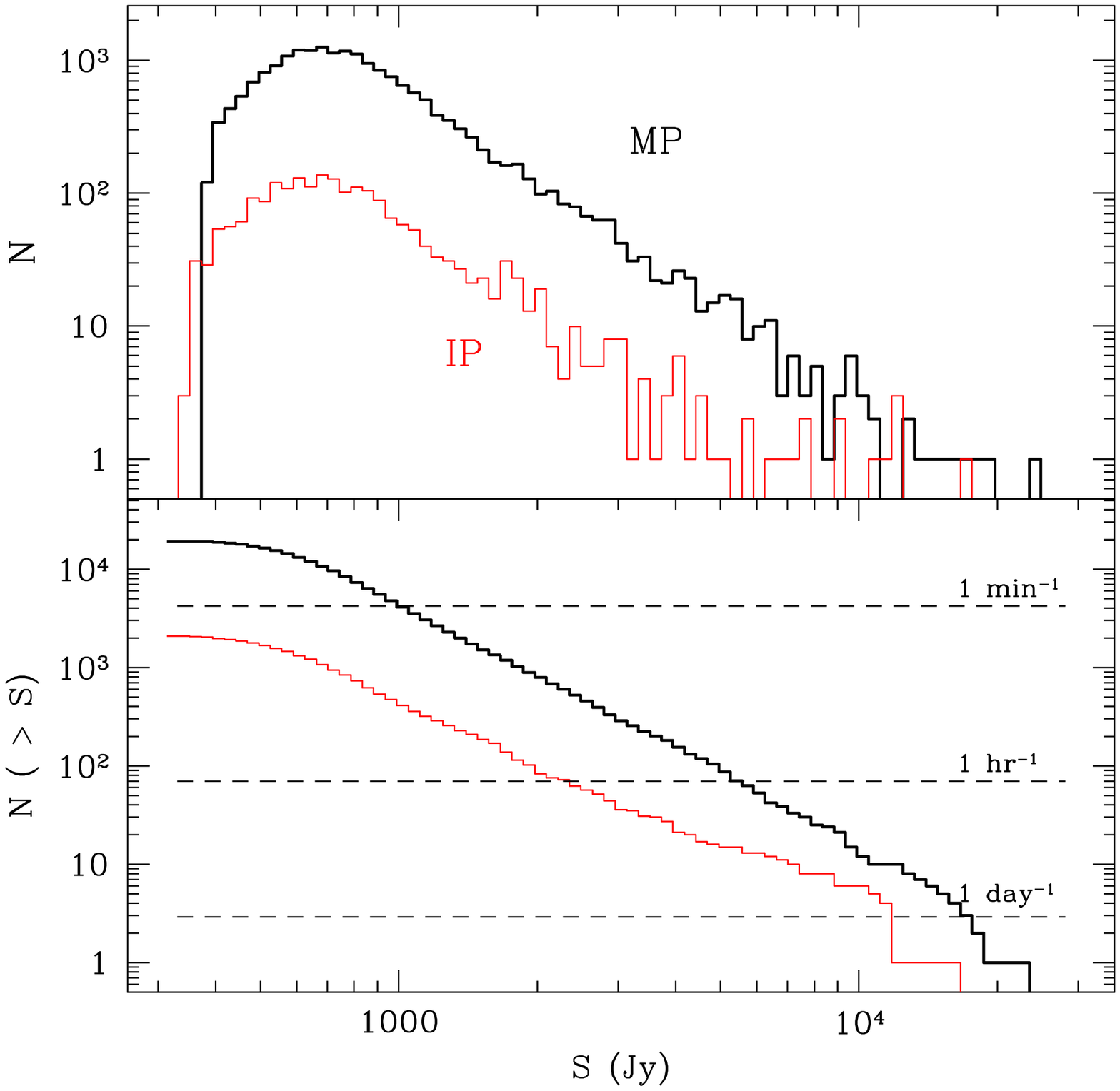}
\figcaption{
\label{fig:ampdist}
Peak flux densities of giant pulses from the Crab pulsar.
The upper panel shows
results obtained from
70 hours of observation at 812~MHz
with the 140-ft telescope at Green Bank \cite{lund95}.
Histograms are shown separately
for the main pulse (MP) and interpulse (IP) components.
The flux scale is different from that in Lundgren et al.
because their Figure 4 gives the flux integrated over a window
that is 1.5 ms wide.  We have converted to peak flux densities
by assuming the typical  pulse width of 100 $\mu$s inferred
by Lundgren et al.  The lower panel shows the corresponding
cumulative distributions, with dashed lines marking
detection rates of one pulse per minute, one pulse per hour and one
pulse per day.
}
\bigskip

\section{Detectability of Giant Pulses} \label{sec:eg}

The vast majority of the over 1500 radio pulsars known were discovered
through searches for periodic signals using Fourier methods. Because these
searches incorporate many pulses from a single source, they are generally
more sensitive than single-pulse searches. However, there are some
conditions under which a single-pulse search may be more sensitive than a
search for periodicity. In this section, we describe how searches for
single pulses may be more effective than periodicity searches at detecting
distant, giant-pulse emitting pulsars.

Finding distant pulsars in other galaxies is important as, of the roughly
1500 radio pulsars known, only $\sim$~25 are extragalactic, all located in
the Magellanic Clouds (A.\ G.\ Lyne, personal communication). Detection of
extragalactic pulsars would allow us to study pulsar populations in other
galaxies and to probe the interstellar medium in those galaxies as well as
the intervening intergalactic medium. Finding more young pulsars like the
Crab is crucial for understanding neutron star formation, supernovae and
the early stages of pulsar evolution. However, Crab-like pulsars in our
Galaxy are difficult to detect due to propagation processes, particularly
scattering, through the Galactic plane. Searching for extragalactic
pulsars is attractive because such searches may not suffer as greatly from
such effects. We show how giant-pulse searches may be the most effective
way of detecting these distant pulsars and use the properties of the Crab
pulsar to infer to what distances we would be able to detect similar
pulsars in other galaxies. We also compare the efficacy of searches for
single pulses with that of traditional periodicity searches.

From the radiometer equation, we calculate the maximum 
distance to which a giant pulse like those observed from the Crab pulsar 
could be detected.
Following the notation of \S3.4 of Paper I and assuming matched
filtering to a pulse that is temporally resolved, the maximum
detectable distance scales with the Crab Nebula's distance
$D_C = 2$ kpc as
\be
{D}_{\rm max} &=& 
	\tsp^{-1/2} 
	D_C 
	\left ( \frac{S_{\rm GP}}{S_{\rm sys}} \right)^{1/2}
	\left( N_{\rm pol} \Delta\nu W\right)^{1/4} \nonumber \\
&\approx& 0.85 \hspace{0.03in} {\rm Mpc}
\left(\frac{\tsp}{5}\right)^{-\frac{1}{2}}
\left(\frac{S_{\rm sys}}{5 \, {\rm Jy^{-1}}}\right)^{-\frac{1}{2}}
\left(\frac{S_{\rm GP}}{10^5 \, {\rm Jy}}\right)^{\frac{1}{2}}
	\nonumber \\
&&\times \left[
		\left(\frac{\Delta\nu}{10 \hspace{0.03in} {\rm MHz}}\right)
		\left(\frac{W}{0.1 \hspace{0.03in} {\rm ms}}\right)
		\left(\frac{N_{\rm pol}}{2}\right)\right]^{\frac{1}{4}}
\label{eq:maxdist}
\ee
where $\tsp$ is the S/N threshold, $S_{\rm sys} = T_{\rm sys}/G$ 
is the system noise in Jansky units (system temperature divided by the
telescope gain), $\Delta \nu$ is the receiver bandwidth,
$W$ is the pulse width, assumed larger than the 
post-detection time constant, or sampling interval, $\tau$, $N_{\rm pol}$ is the number of
independent polarization channels summed and $S_{\rm max}$ is the 
peak flux density of the giant pulse in the time of observation.
We have assumed in deriving Eq.~\ref{eq:maxdist} that pulses are
dedispersed and that any residual pulse smearing (e.g. from scattering) is
less than $W$. Using fiducial values for an Arecibo 430-MHz search
($\Delta\nu = 10$ MHz, $N_{\rm pol} = 2$, $W \sim 0.1$~ms, 
$S_{\rm sys} = 5$ Jy), and a threshold $\tsp$ = 5, we calculate
$D_{\rm max} \sim$~0.85~Mpc for Crab-like pulsars in a one-hour observation (i.e. $S_{\rm max} \sim
10^{5}$~Jy).
For millisecond pulsars like PSR~B1937+21 ($D = 3.6$ kpc)  and
PSR~B1821$-$24 ($D \approx 5.5$ kpc)  
for which $S_{\rm max} \sim 10^{3}$~Jy in one hour, 
$D_{\rm max} \sim 0.15$ to 0.23 Mpc. 

Given the distribution of giant-pulse intensities measured by
Lundgren et al. (1995), we calculate the number of Crab-like giant pulses
detectable out to a distance $D_{\rm max}$ in such a search. As shown in
Figure~\ref{fig:crabdist}, Crab-like pulsars within several irregular and
spiral galaxies in the Local Group, at distances less than 1 Mpc, would be
detectable at rates greater than one pulse per hour.  These include the
LMC, SMC, M31, M33, NGC6822 and IC1613. If giant pulses an order of
magnitude greater than the Crab's are emitted, then another 7 galaxies in
the 2 to 4.3 Mpc range, including M81, M82, M83, M94, M63, M51 and M101,
might be accessible. Making the naive assumption that all pulsars of the
Crab's age ($\sim$~1000~yr) or less will emit giant pulses, and assuming a
pulsar birthrate of 1 per 100 years, we estimate that at least 10
giant-pulse emitting pulsars should be within each spiral galaxy similar
to our own.
Searching for these young pulsars is attractive as such
searches may not be as impeded by dispersion and scattering as are
searches for pulsars within our own galaxy.

\medskip
\epsfxsize=9truecm
\epsfbox{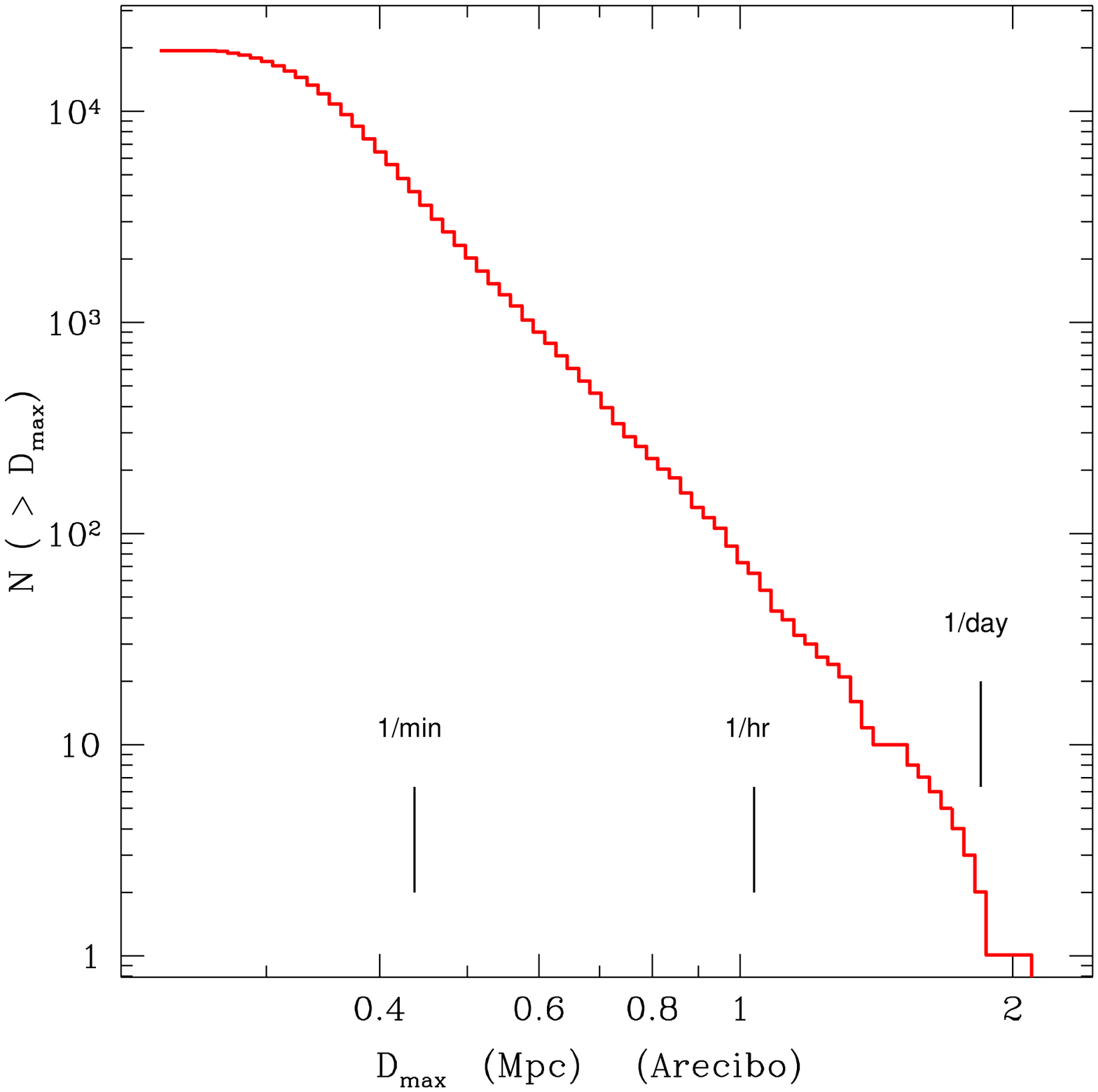}
\figcaption{
Number of detectable Crab-like giant pulses out to a distance
D$_{\rm max}$ for a 70-hr Arecibo search at 430 MHz
with bandwidth of 10 MHz,
a 5$\sigma$ detection threshold and matched filtering
(c.f. Eq.~\ref{eq:maxdist}).
We have scaled the distribution of pulse amplitudes at 812 MHz from
Lundgren et al. (1995) by assuming a spectral index of
$-4.5$ for giant pulses.  This spectral index is consistent with
the work of Sallmen et al. (1999). Also, the predicted
rate of detections is consistent with recent Crab pulsar
observations we have made (Cordes et al. 2003)
at 430 MHz which yield
$S_{\rm max} = 1.6 \times 10^5$ Jy in one hour of observing
and imply $D_{\rm max} = 1.6$ Mpc.
\label{fig:crabdist}
}
\bigskip

It is useful to compare single-pulse and periodicity searches 
and the resultant maximum distances to which an isolated pulsar would be 
detectable through the two methods. Letting
D$_{\rm max,SP}$ and D$_{\rm max, FFT}$ be the maximum distances to which
we could detect a pulsar through a search for single pulses and through a
FFT-based standard periodicity search, their ratio,
\begin{equation}
 \frac{{\rm D}_{\rm max,SP}}{{\rm D}_{\rm max,FFT}} =
\left(\frac{\tsp({\rm FFT})}{\tsp({\rm SP)}}\right)^{\frac{1}{2}}
 \left(\frac{S_{\rm max}}{S_{\rm avg}}\right)^{\frac{1}{2}}
\left(\frac{\tau}{T}\right)^{\frac{1}{4}}
\left(\frac{W}{P}\right)^{\frac{1}{4}},
\label{eq:snrratio}
\end{equation}
depends on the thresholds for the periodicity and single-pulse searches
$\tsp({\rm FFT})$ and $\tsp({\rm SP})$, the post-detection time constant $\tau$
(optimally equal to the pulse width $W$), the total time integrated $T$, the
maximum pulse giant pulse flux density, $S_{\rm max}$, 
seen in that time, the average flux
$S_{\rm avg}$, and the period of the pulsar $P$.  
For typical parameters 
($\tsp({\rm SP})$ = 5, 
$\tsp({\rm FFT})$ = 8, 
$T$ = 3600 s,
$\tau$ = 0.1~ms 
and $W/P$ = 0.05), 
we find that a search for single pulses
can detect pulsars to greater distances than the periodicity search when
$S_{\rm max}/S_{\rm avg} > 1.7\times10^{4}$. 
The Crab pulsar, with $S_{\rm max}/S_{\rm avg} = 1.5\times10^{5}$ 
satisfies this criterion while PSR~B1937+21, with 
$S_{\rm max}/S_{\rm avg} = 4.2\times10^{3}$, does not.

A more detailed analysis in the Appendix compares the
two methods  for different pulse-amplitude distributions.
For distributions that are strictly power-law in form and with
a large range of pulse amplitudes, single pulse searches
are superior to periodicity searches 
for power-law indices  $\alpha$ in the range of 
$1 \lesssim \alpha \lesssim 3$ when the number
of pulse periods investigated ranges from $\sim 10$ to $10^5$.
For the Crab pulsar,  $\alpha \approx 3.5$, but single-pulse
detection is superior to a periodicity search because pulses
comprise both a power-law distribution and 
a component of much weaker but more frequent pulses, 
We find that in some
circumstances, single-pulse searches are more effective than searches for
periodicity, {\it even in the absence of giant pulses}. These regimes are
further outlined in the Appendix and show that pulsar searches should
always incorporate single-pulse searches in addition to standard
periodicity searches. The computational time needed for the single-pulse search is typically a
small fraction ($< 10\%$) of the total search time. We note that the relative sensitivites of
single-pulse and periodicity searches will depend on the extent and nature of radio
frequency interference (RFI), with single-pulse searches being more affected by bursty, aperiodic RFI.

\section{Searches for Extragalactic Pulsars} \label{sec:targets}

We now describe the results of searches for giant pulses from the nearby
spiral galaxy M33, PSR~B0540$-$69 in the LMC and from several other
galaxies. 
While these searches were designed to detect giant pulses from
pulsars, they are sensitive to other transient radio sources and serve as
practical examples of applying the methodology described in Paper I.
They also illustrate the challenges of searches for fast radio transients
in the presence of RFI.

\subsection{M33} \label{sec:M33}

As shown in Figure~\ref{fig:crabdist}, all of the galaxies in the Local
Group are within the range of detection of Crab-like giant pulses. The
spiral galaxy M33, at a distance of approximately 840 kpc
\cite{freedman91,freedman2001}, is an excellent candidate for giant-pulse
detection because its light derives mainly from blue, supergiant stars
(i.e. progenitors of neutron stars). It is also visible with Arecibo,
currently the most sensitive instrument for single-pulse searches at
$\sim$~meter wavelengths. Our Arecibo search used a radio frequency of 430
MHz, optimal because of the steep spectral indices of pulsars and the low
dispersion and scattering expected towards M33. The entire galaxy was 
covered with $16\times9'$-diameter beams, as shown in
Figure~\ref{fig:m33beams}. In Table~\ref{tab:m33beams} we list beam number, RA and DEC
in J2000 coordinates, the number of pointings at each beam and the total observation time per beam.
Each beam was observed five to ten times for
900 or 1800 seconds over MJDs from 51054 to 52406 (i.e. from 29 August
1998 to 12 May 2002). The total amount of time spent per beam over these
dates ranged from 1.75 to 3.0 hours. Two off-source positions were also observed,
to provide an independent test of the reality of any detected signals.
Observations were done with the
Arecibo Observatory Fourier Transform Machine (AOFTM,
\verb+http://www.naic.edu/~aoftm/+), which covers a 10-MHz bandwidth with
1024 frequency channels and uses a sampling time of 102.4~$\mu$s.

\medskip
\epsfxsize=9truecm
\epsfbox{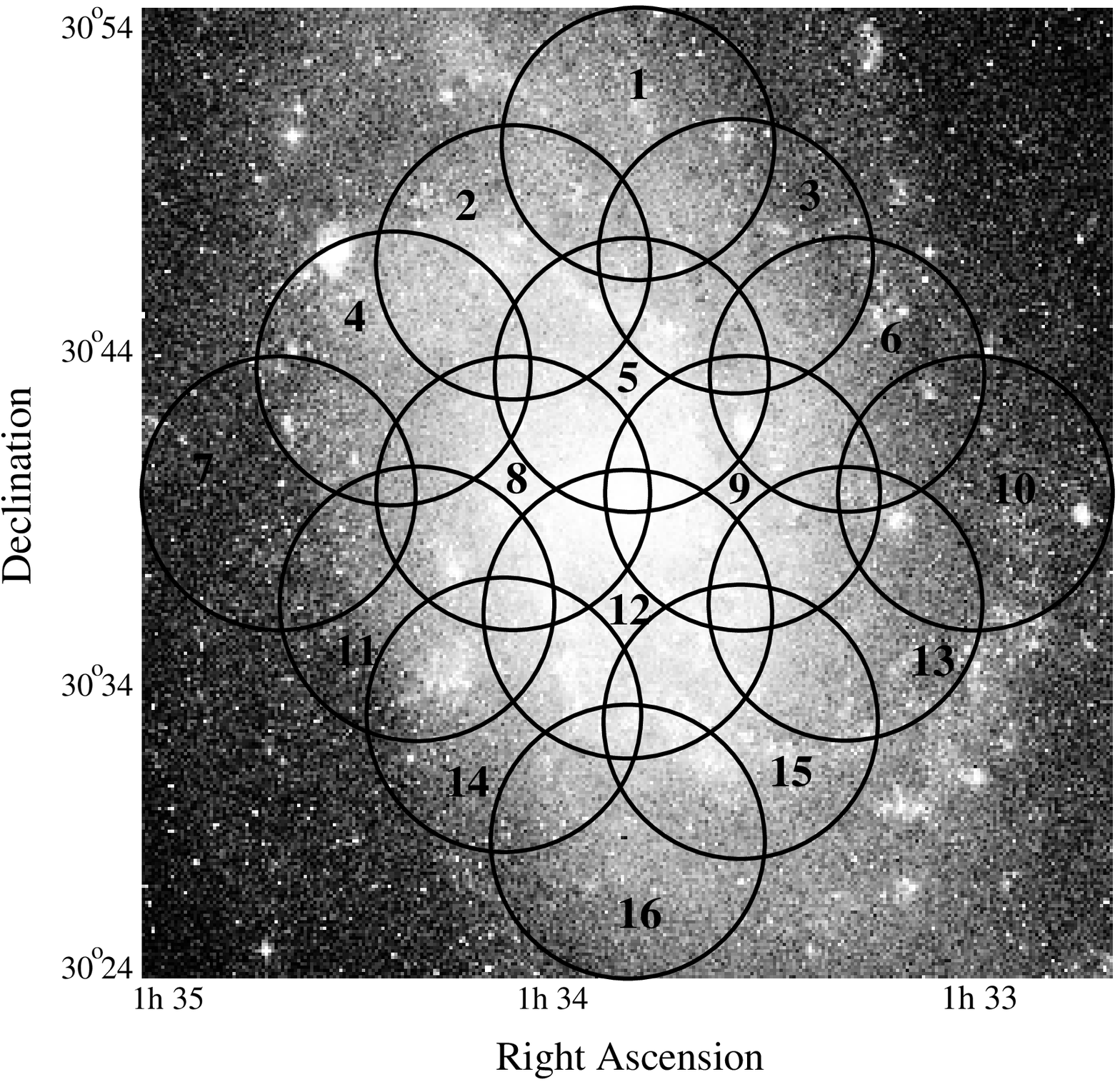}
\figcaption{
The sixteen beams used for our Arecibo giant-pulse search superposed on an
image of M33 from the Digitized Sky Survey \cite{digit}. Gordon et al
(1999) have cataloged 53 radio-bright supernova remnants in M33, the
largest extragalactic sample ever. These remnants are fairly uniformly
spaced out to a distance of roughly 15 arcminutes from the center of the
galaxy, with the most distant remnant lying approximately 20 arcminutes
from the center. There are a total of 98 cataloged supernova remnants in M33, compared
with 231 in our Galaxy \cite{green02}.
 \label{fig:m33beams}
}
\bigskip

Due to the high latitude ($|b| \sim 30^{o}$) of M33, the predicted
Galactic contribution to the dispersion measure (DM) along the line of sight is only $\sim$ 50
pc cm$^{-3}$ \cite{cl02}.  As the orientation of M33 is nearly face-on
\cite{garcia91}, we take Galactic and extragalactic DM contributions to be
equal and estimate the maximum dispersion smearing expected across a
10-kHz frequency channel to be $\sim$~0.1~ms. To allow for errors in the
model and any regions of enhanced electron density along the line of
sight, our searched DMs ranged from 0 pc cm$^{-3}$ to 250 pc cm$^{-3}$.
Scaling from the Green Bank measurements of Lundgren et al. (1995), as in
Figure~\ref{fig:crabdist}, we estimate that more than one pulse per hour 
should be detectable from a Crab-like pulsar in M33 above a S/N threshold
of 5, corresponding to a threshold of 0.5 Jy. Assuming a pulsar
birthrate of 0.01 yr$^{-1}$, we expect there to be roughly 10
pulsars of the Crab's age or less in our Galaxy.
Because the mass of M33 is roughly 10 times less than that of our
Galaxy \cite{mendez99,corbelli00,sakamoto03}, we might expect 10 times fewer young,
Crab-like pulsars. However, evidence suggests that the supernova rate per unit mass is
higher in M33 than in our Galaxy \cite{berk84,gordon98}. Because
these mass and supernova rate
estimates are highly uncertain, it is difficult to predict how many Crab-like pulsars
there might be in M33, but we expect it to be of order $\sim$ 10. 
Furthermore, since giant pulses are also emitted by millisecond
pulsars like PSRs~B1821$-$24 and B1937+21, the number of detectable pulsars may be larger.

Because RFI contaminated much of our data, 
our actual threshold was often significantly higher than
that calculated from radiometer noise alone. 
In some cases, RFI was minor and
did not greatly influence our sensitivity to pulses of astrophysical
origin. In other cases, however, the RFI was quite strong and significant
portions of the data had to be discarded from the analysis.
This was done with interactive tools for displaying and selecting subsets of
data in time and DM space.

For each beam, we applied the search algorithms described in Paper I to
create diagnostic plots like those shown in Figures~\ref{fig:m33giant1},
\ref{fig:m33giant2} and \ref{fig:m33giant3} and inspected the results 
for evidence of high-DM pulses of an astrophysical origin. In Paper I
similar plots for the giant-pulsing pulsars PSRs~B0531+21 and B1937+21
are shown.  For our M33 search, a signal-to-noise threshold of 4 was used
and each time series was smoothed in seven stages (where adjacent samples
were summed in each stage), corresponding to a maximum time
resolution of $2^7\times 102.4\,\mu s\approx $~13~ms. 
If a pulse is detected in several of 
the smoothed time series, 
only the highest S/N pulse is recorded. 
Most beams,
like that shown in Figure~\ref{fig:m33giant1}, 
show an excess of pulses at low DM due to RFI but show no evidence 
for high DM pulses. For some beams,
the most dramatic example of which is shown in Figure~\ref{fig:m33giant2},
we detect an excess number of pulses at a well-defined, non-zero DM. For
other beams, as shown in Figure~\ref{fig:m33giant3}, we detect individual
strong pulses at high DM. While these features are tantalizing and, in
some cases, consistent with originating from a pulsar in M33, no features
repeat on multiple pointings at the same position. However, this is not inconsistent with
the predictions of Figure~\ref{fig:crabdist}.
 For each beam, we summed all of the DM histograms (i.e. upper
right panels of Figures~\ref{fig:m33giant1}, \ref{fig:m33giant2} and
\ref{fig:m33giant3}) to determine if low-level excesses of pulses at a
specific DM were present. However, no features were apparent in these
combined histograms. 

It is unlikely that interstellar scintillation is responsible for 
features, such as those shown in Figures~\ref{fig:m33giant2} 
and \ref{fig:m33giant3},
appearing only sporadically. 
Assuming a Galactic scale height of 1~kpc for 
$C_n^2$ (the electron-density fluctuation spectral coefficient;
see Paper I) and a source velocity of 
200~km/s (e.g. Arzoumanian et al. 2002) and applying the results of the
Appendix of Paper I, we find that a source in M33
should have a scintillation timescale of only $\sim$~35~seconds and a
scintillation bandwidth of only $\sim$~20~kHz at 430~MHz, rendering ISS
unimportant in modulating source intensities during or between our
observations.  At 1400~MHz, these timescales and bandwidths would be much
larger (roughly 141~seconds and 3.6~MHz) and would be important to consider.

Unfortunately, for the time being, our search for giant pulses from M33
pulsars remains inconclusive. It could be that the Crab and PSR~B0540$-$69 are
unusual pulsars and that there are no similar, giant-pulse emitting
pulsars in M33. It is also possible that we have detected giant pulses
from M33 pulsars, but they are either too sporadic or too weak to be
distinguished from the many pulses due to RFI. Sensitive, multiple-beam
and/or multiple-site observations are necessary to conclusively 
assess the reality of the signals we have detected.

\medskip
\epsfxsize=9truecm
\epsfbox{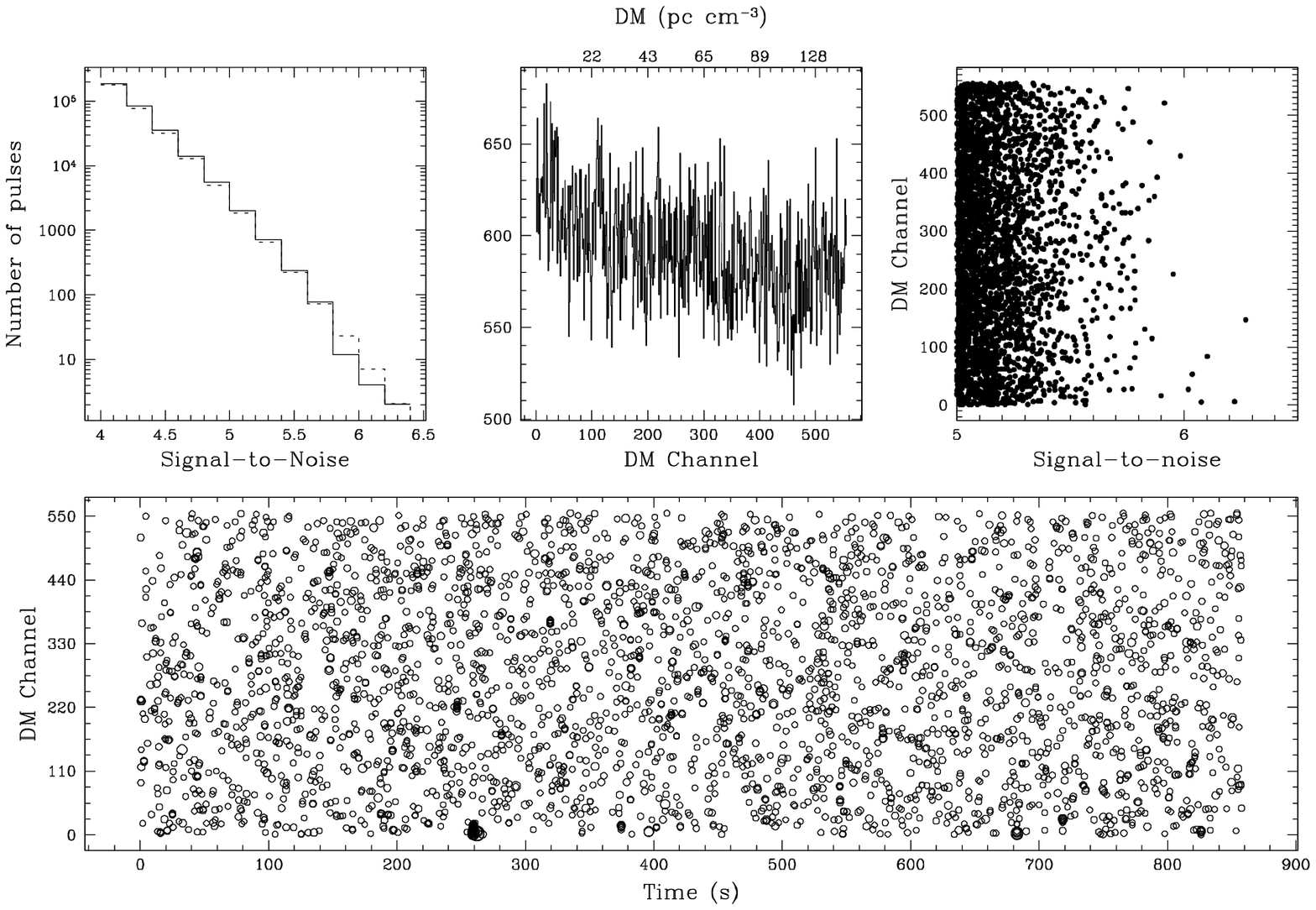}
\figcaption{
Single-pulse search results for M33 beam 4 on MJD 52405. {\sc Upper
left:} The number of pulses above a 4$\sigma$ threshold vs. signal-to-noise. Note the
logarithmic scale of the y-axis. The dotted line shows the expected
distribution for noise only, as describedin Paper I. {\sc Upper center:} Number of pulses above
a 4$\sigma$ threshold vs. DM channel. The lower x-axis shows DM channel
number, while the upper x-axis shows DM in pc cm$^{-3}$. We have used the
optimal DM channel spacing described in Section 3.2 of Paper I. {\sc Upper right:}
DM channel vs signal-to-noise for all pulses with S/N greater than 5$\sigma$.
To create these three upper plots, pulses which are strongest at zero DM have been excised. {\sc
Bottom:} All pulses with signal-to-noises greater than 5$\sigma$ plotted
 vs. DM channel and time. The size of the circle is linearly proportional
to the signal-to-noise of the pulse. Even though we ignore them in our analysis, pulses which
are strongest at low DM are shown in this bottom plot to illustrate the effects of RFI.
 As for the majority of M33 observations, there are no strong individual pulses at high DM.
Likewise, an excess of pulses at high DM in the DM histogram is not
apparent.
\label{fig:m33giant1}
}
\bigskip

In addition to the single-pulse search, a periodicity search was performed
on all of the M33 data, typically using 8-Mpoint FFTs (i.e. roughly 14
minutes of data at our 102.4~$\mu$s sampling rate). While the chances of
detecting a pulsar in M33 through such a search are very small (the pulsar
would have to have an average flux density of approximately 30 times that
of the Crab's), we were sensitive to pulsars within our own Galaxy along
the line of sight to M33. No periodic sources were detected in this search
down to a limiting sensitivity of 0.2~mJy (assuming a duty cycle of 0.05).

\medskip
\epsfxsize=9truecm
\epsfbox{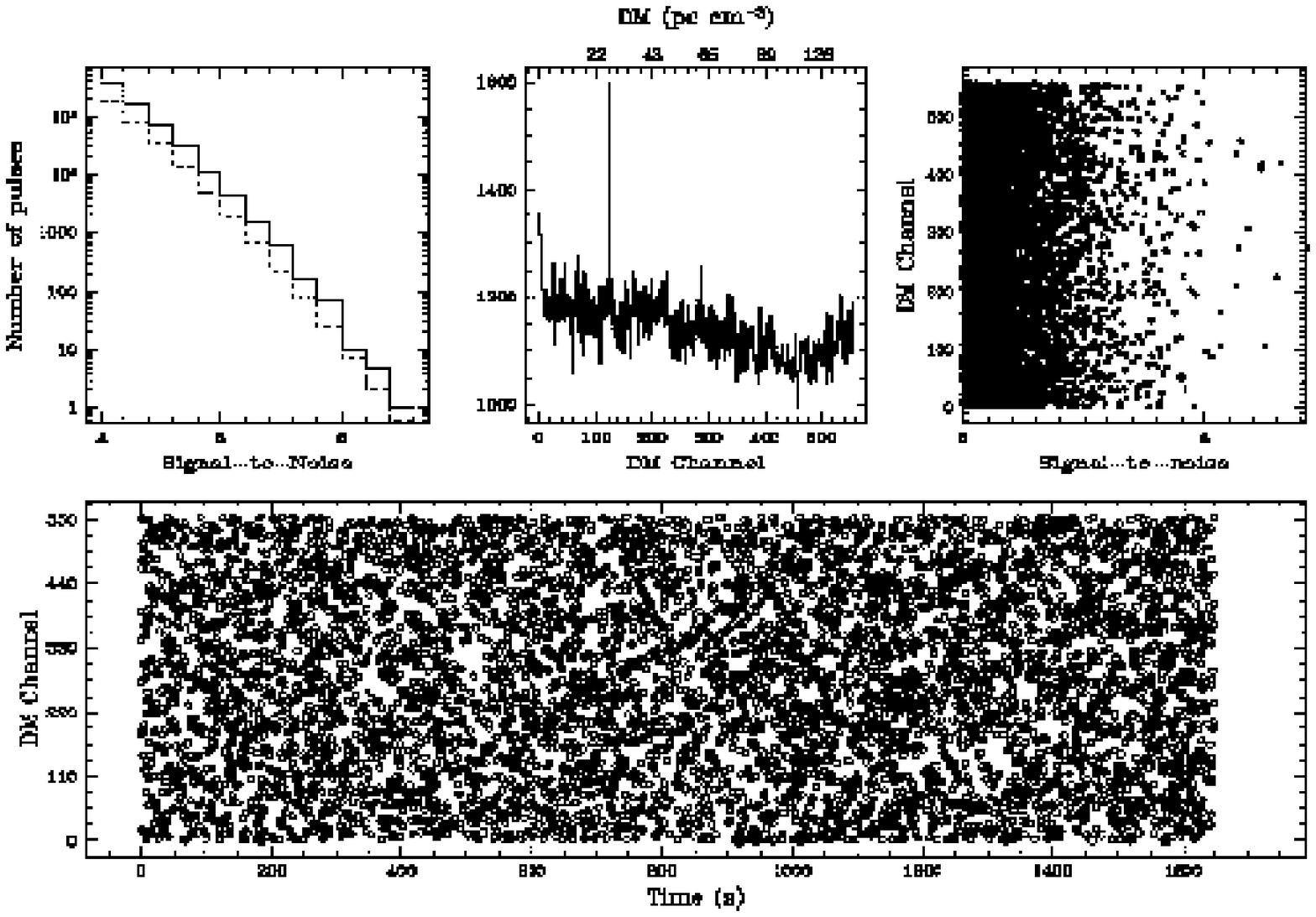}
\figcaption{
Single-pulse search results for M33 beam 13 on MJD 51918. An excess of single pulses at
DM $\sim$ 26.5 pc cm$^{-3}$ is obvious. This excess is due to many weak pulses; no strong
single pulses at this DM are seen. This pulse must have
width $<$ 0.1 ms to show such a narrow DM distribution (see Figures 3 and 4 of Paper I).
\label{fig:m33giant2}
}
\bigskip

\medskip
\epsfxsize=9truecm
\epsfbox{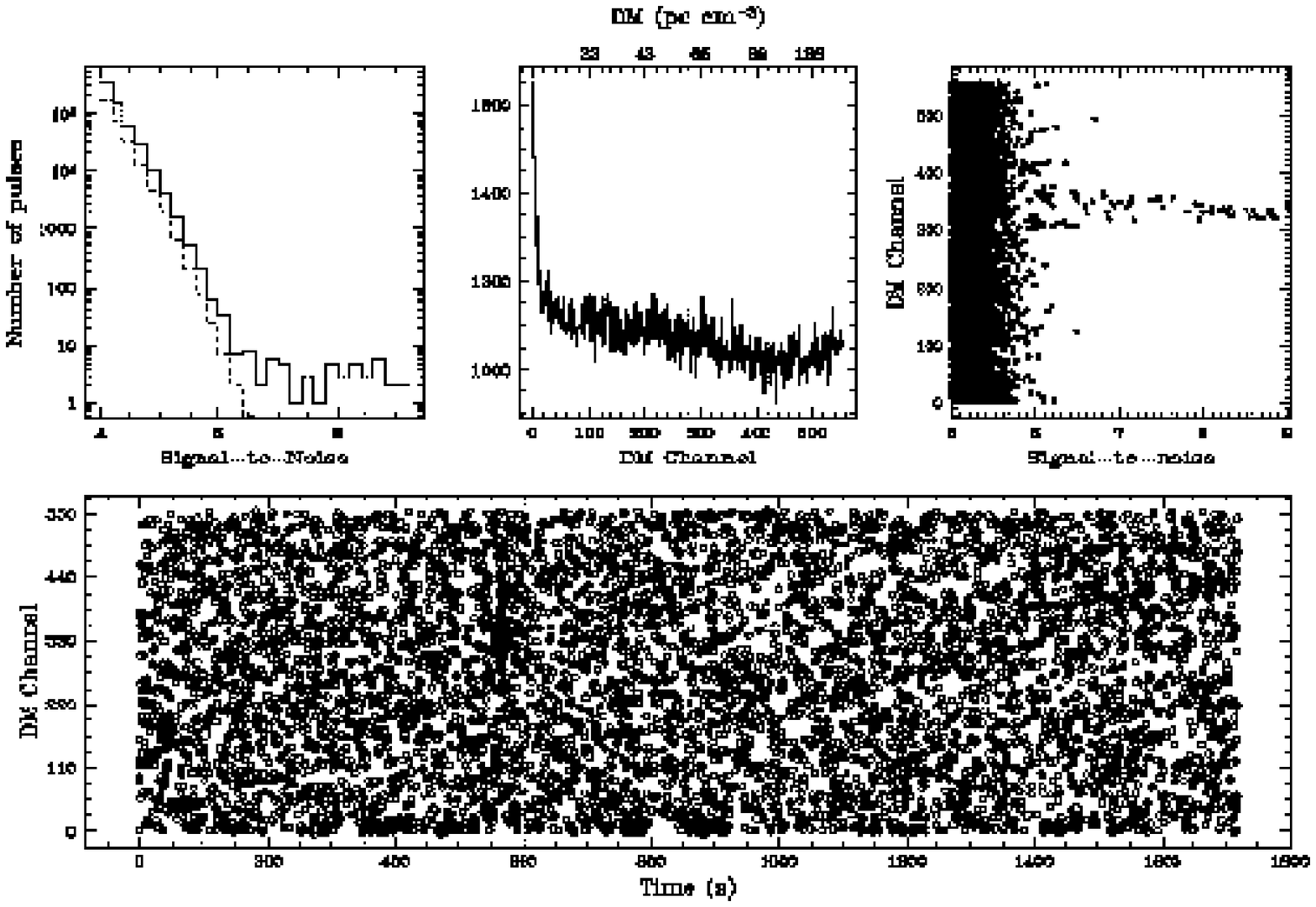}
\figcaption{
Single-pulse search results for M33 beam 16 on MJD 51921. In addition to
an excess of pulses at zero DM due to RFI, a single pulse with
signal-to-noise $\sim$ 9 is detected. The DM of this pulse is $\sim$ 71 pc
cm$^{-3}$, consistent with an origin in M33.  This width of this pulse is
approximately 1~ms.
\label{fig:m33giant3}
}
\bigskip

\medskip
\epsfxsize=9truecm
\epsfbox{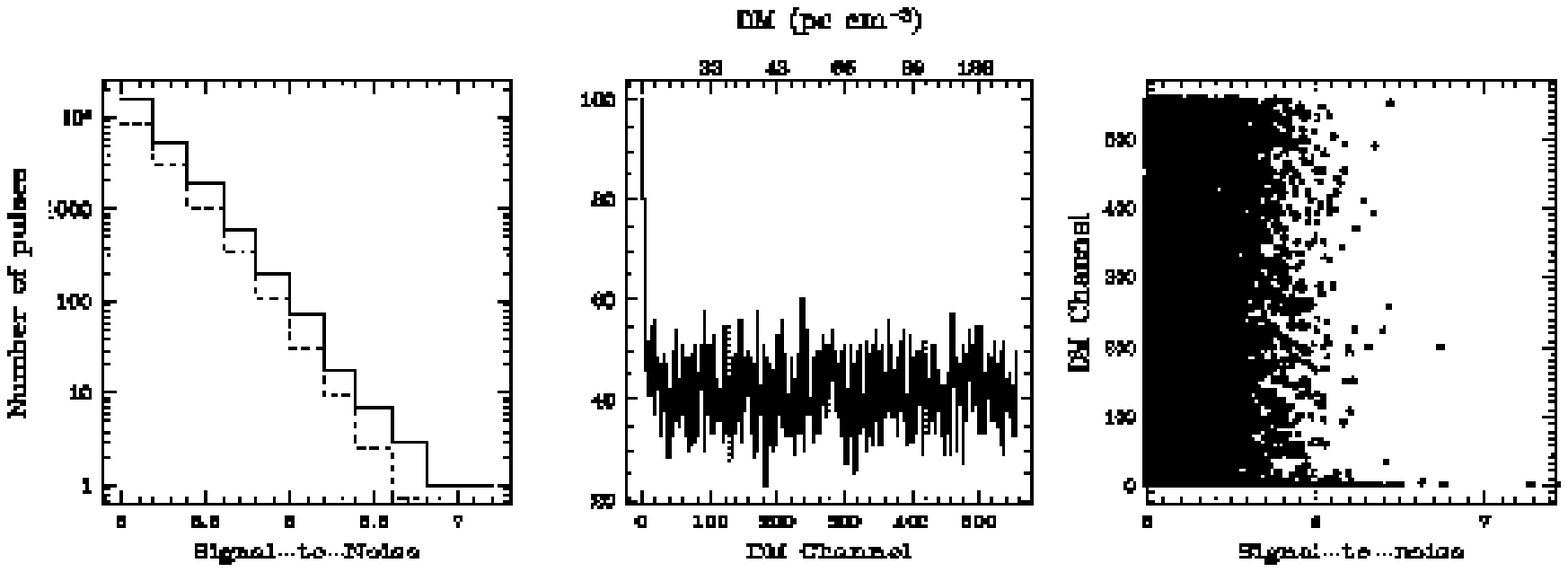}
\figcaption{
Single-pulse search results for pointings away from M33. These plots show cumulative results
for all pulses above 5$\sigma$ for all five 1800-second off-source datasets. No strong individual pulses
or excesses in the DM histogram are detected.
\label{fig:m33off}
}
\bigskip

\subsection{PSR~B0540$-$69} \label{sec:0540}

With a period of 50.3~ms, a period derivative of $4.8\times10^{-13}$ s
s$^{-1}$, and hence a spin-down age of roughly 1660 years, pulsar
B0540$-$69 has more similar spin-down parameters to the Crab pulsar than
any other known pulsar. Furthermore, as shown in Table~\ref{tab:blcs}, the
magnetic field at the light cylinder of this pulsar, the strength of which
seems to be correlated with giant-pulse emission, is fifth among known
radio pulsars. Although it is roughly 50~kpc away in the LMC, it is a
strong X-ray and optical source. The DM of PSR~B$0540-69$ measured by
Manchester et al. (1993) is $146 \pm 4$ pc cm$^{-3}$. They measure a
640~MHz flux of 0.4~mJy, roughly the same flux that the Crab would have at
the distance of the LMC. While its integrated radio emission is weak, we
might expect PSR~B0540$-$69 to emit giant pulses similar to those detected
from the Crab. The detection of such pulses from PSR~B0540$-$69 would
allow us to compare the shapes and phases of giant pulses with those at high energies,
determine the dispersion measure more accurately and
perhaps measure polarization of strong giant pulses.

With this in mind, we obtained a total of 55 observations of
PSR~B0540$-$69 between September 7, 1994 and September 26, 1994 with the
Parkes Radiotelescope in New South Wales, Australia. These include
15 observations at a center frequency of
660~MHz, with 256~0.125-MHz channels covering a 32-MHz bandpass, and 40
observations at a center frequency of 1520~MHz, with 64~5-MHz channels
covering a 320-MHz bandpass. At both
frequencies, a sampling time of 0.6~ms was used. Most of our pointings
were one-hr long, bringing the total number of hours on source to
14.4~hours at 660~MHz and 38.1~hours at 1520~MHz. Our 5$\sigma$
single-pulse detection thresholds calculated from radiometer noise were
4~Jy and 0.5~Jy at 660 and 1520~MHz, respectively. Actual thresholds were
significantly higher in some pointings due to RFI, but most RFI could be
excised by discarding subsets of the data from the analysis.

Given the parameters above, we would expect to detect Crab-like giant
pulses up to S/Ns of 9 and 17 at the lower and upper frequencies,
respectively, in our total observation times at the two frequencies (assuming a spectral index of
$-2$). In Figures~\ref{fig:0540_660} and \ref{fig:0540_1520}, we plot the
signal-to-noises and DMs of all pulses
detected above a 5$\sigma$ threshold at both frequencies. 
Our data at 660 MHz do not show any evidence for giant pulses
at the DM of B0540$-$69. 
 At 1520 MHz, the strongest non-zero DM pulse has a
amplitude of 7$\sigma$ and a DM of 146~pc~cm$^{-3}$. While this
is tantalizing, several pulsars nearly as strong are detected at other DMs, and 
we cannot claim detection of giant pulses with our data.
Given the large scattering timescales measured by JR03, it is unlikely
that scintillation played a role in modulating the fluxes of pulses during
our observation.

\medskip
\epsfxsize=9truecm
\epsfbox{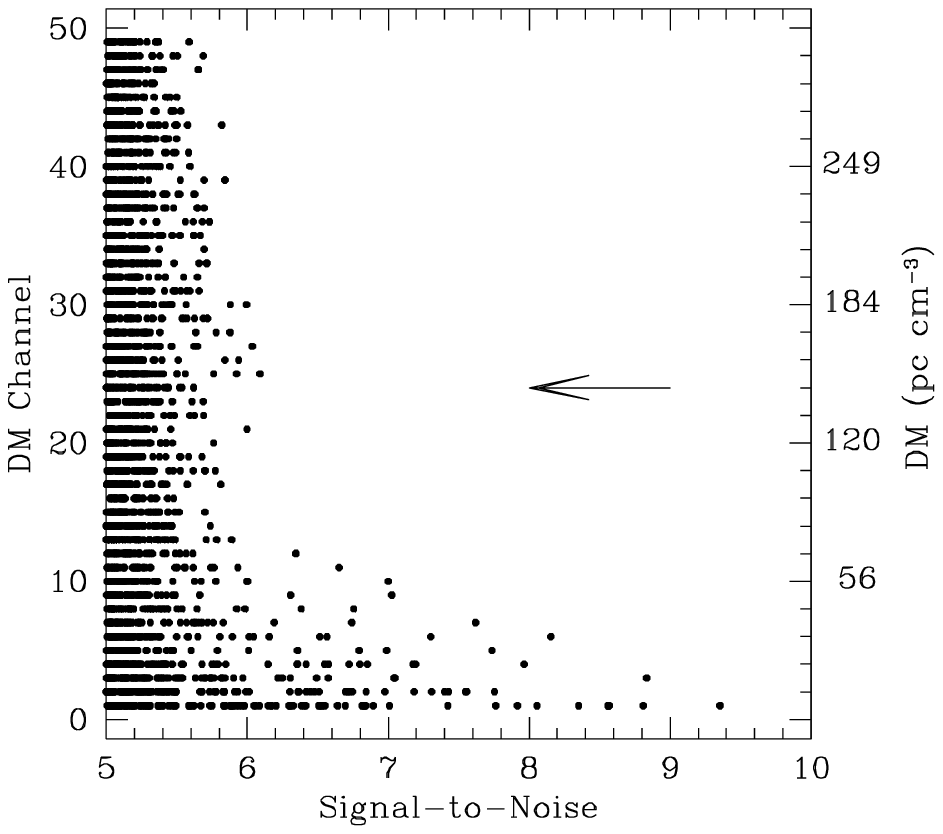}
\figcaption{
DM vs. signal-to-noise for all pulses detected above a threshold of 5$\sigma$
from PSR~B0540$-$69 at a center frequency of 660~MHz. The arrow shows the pulsar's DM
of 146~pc~cm$^{-3}$.
\label{fig:0540_660}
}

\bigskip

\medskip
\epsfxsize=9truecm
\epsfbox{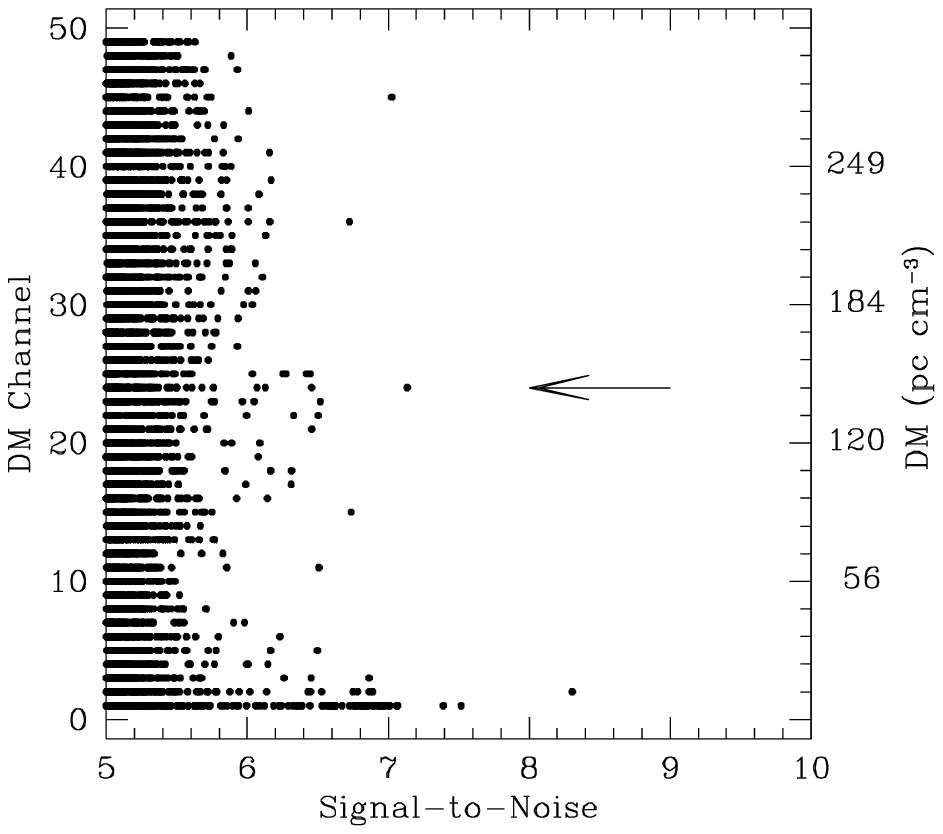}
\figcaption{
DM vs. signal-to-noise for all pulses detected above a threshold of 5$\sigma$
from PSR~B0540$-$69 at a center frequency of 1520~MHz. The arrow shows the pulsar's DM
of 146~pc~cm$^{-3}$.
\label{fig:0540_1520}
}

\bigskip

While the 1384-MHz JR03 observations benefited from a new Parkes receiver with
lower system temperature and
a slightly lower observing frequency, the main difference between their observations
and the observations reported here is the channel bandwidth. They used 0.5-MHz filters,
implying smearing of only 230~$\mu$s for a DM of 146~pc~cm$^{-3}$. Our 5-MHz
filters caused a DM smearing of 1.8~ms, much more than the scatter-broadened giant-pulse width
of 0.4~ms measured by
JR03.   
Given the peak giant-pulse fluxes and widths reported by JR03,
we would expect to detect their strongest pulse at a
signal-to-noise of $8\sigma$. Adding in the effects of a slightly higher observing frequency
and increased system noise, the pulse signal-to-noise distribution in Figure~\ref{fig:0540_1520} 
is consistent with their observations.

Our non-detection of giant pulses at 660 MHz is more difficult to understand.
Given the 0.4-ms scatter-broadened width of the pulses detected by JR03, we expect 10-ms of
scatter broadening at 660~MHz, much greater than the 0.5-ms expected DM smearing across one
of our 0.125-MHz
channels. If the spectral index of  PSR~B0540$-$69 is indeed as steep as $-4.4$,
as the observations by JR03 suggest, we would expect to detect pulses as strong as 20$\sigma$
in our data.
A spectral index of $-3$ or higher is required to explain our null results at 660 MHz.

\subsection{Other Galaxies} \label{sec:other}

During our search for giant pulses from PSR~B0540$-$69, we observed
several other galaxies. While it is unlikely that a pulsar could be
detected in any of these galaxies, aside from the LMC, these searches are
constraining for some of the other source populations discussed in Section
2 of Paper I.
 In Table~\ref{tab:othergal}, we list these
targets, along with Right Ascension, Declination, Galactic longitude,
Galactic latitude, estimated distance and the number of pointings $N_{p}$
at each object. Each pointing was 1800-seconds long and for most pointings we used
a central frequency of 435~MHz, with 256 channels of width 0.125-MHz and a sampling time
of 0.42~ms.
Galaxies NGC253, NGC300, Fornax, NGC6300 and NGC7793 were
all covered by a single Parkes beam, while 6 separate pointings were
necessary to cover the 2~deg$^{2}$ error box of the high-energy LMC source
3EG~J0533$-$69 \cite{hartman99}. We performed a search for single pulses on all of the
pointings, but did not find any conclusive evidence for isolated dispersed pulses
from any of these galaxies. In Figure~\ref{fig:0537_giant}, we present results for
the LMC pointing with the 16-ms X-ray pulsar J0537$-$6910 in the beam. This pulsar is
interesting as it has a $B_{lc}$ which is over twice that of the Crab or PSR~B1937+21.
We detect an excess of pulses at a DM of 113~pc~cm$^{-3}$,
consistent with the DMs of pulsars already detected in the LMC \cite{crawford01}.
However, because of the presence of RFI which manifests itself as dispersed radio
pulses at many DMs, it is impossible to assess the reality of this signature.
More sensitive and/or multiple beam/multiple site observations are required.

\medskip
\epsfxsize=9truecm
\epsfbox{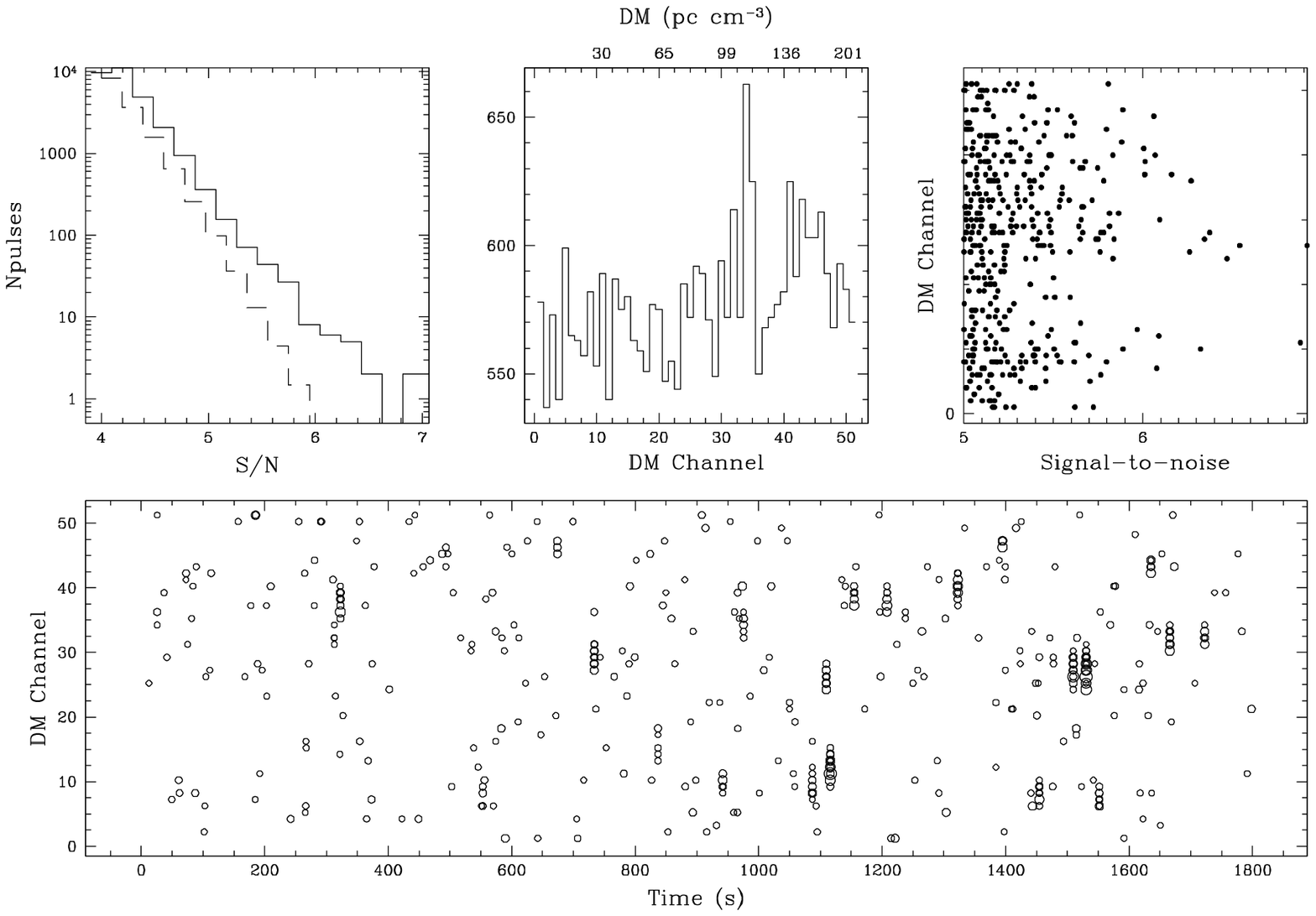}
\figcaption{
Single-pulse search results for the 1800-second LMC pointing at 435~MHz with the
16-ms X-ray pulsar J0537$-$6910
in the beam. While an excess number of pulses is detected at a DM $\sim$ 113 pc~cm$^{-3}$,
strong pulses across many DMs due to RFI make assessing the reality of this excess
difficult.
\label{fig:0537_giant}
}
\bigskip

\section{Conclusions} \label{sec:conclusions}

We have described the capabilities of searches for extragalactic giant
pulses and the results of several searches towards extragalactic targets.
We find that giant pulses from Crab-like pulsars should be detectable from
all galaxies in the Local Group. We explore different pulse-amplitude
power distributions and intensity cutoffs and find that single-pulse
searches are superior to periodicity searches in certain cases. We
emphasize that this kind of search should be done along with standard
periodicity searches for pulsars because of the small additional
computational efforts and the possibly large gains. The example results
from searches for giant pulses from extragalactic pulsars illustrate the
methodology and challenges of any search for transient radio signals.  We
have detected several intriguing dispersed pulses of possibly
astrophysical origin from the spiral galaxy M33 with a search using the
Arecibo radiotelescope at 430~MHz. Assessment of the astrophysical nature
of these pulses must wait for more sensitive multiple-beam and/or
multiple-site observing systems. We have detected no convincing signatures of
giant pulses from
the young LMC pulsar B0540$-$69 or from several other galaxies in searches
with the Parkes radio telescope.

While the results of the M33 search presented here are inconclusive, they
illustrate both the capabilities and difficulties in single-pulse searches
and serve as a pilot study for future searches.  To increase the
capabilities of single-pulse searches, systems with multiple beams or
multiple antennas are necessary in order to perform coincidence tests
which will discriminate between pulses of astrophysical and terrestrial
origins. Larger collecting areas are needed to increase our sensitivity to
pulses from pulsars and other transient sources in other galaxies.
Multi-beam systems for the Arecibo and Green Bank telescopes will likely
be built within the next few years. Observing with multiple beams will
allow us to distinguish RFI from real signals and would remove much of the
ambiguity in interpreting the results of our M33 search. On a longer time
scale, large radio arrays such as the Square Kilometer Array (SKA) will be
developed. With a planned field of view over 100 times that of Arecibo and
a sensitivity over 20 times that of Arecibo, the SKA will revolutionize
single-pulse searches. While the SKA will not be operational for many
years, single-pulse searches in the meantime, especially with multiple
beams, are essential for optimizing routines for RFI excision and for
processing large amounts of data.

\acknowledgments

We thank Simon Johnston and Froney Crawford for useful discussions.
MAM is supported by an NSF Math and Physical Sciences Distinguished
Research Fellowship. Arecibo Observatory is operated by the National
Astronomy and Ionosphere Center, which is operated by Cornell University
under cooperative agreement with the National Science Foundation (NSF).
The Parkes Observatory is part of the Australia Telescope which is funded
by the Commonwealth of Australia for operation as a National Facility
managed by CSIRO.

{}

\clearpage
\appendix

\section{Periodicity Searches vs. Single-Pulse Searches}
\label{app:searches}

\newcommand{\iavephi}{\langle S \rangle_{\phi}}
\newcommand{\hsum}{h_{_\Sigma}}
\newcommand{\snrpkfactor}{(\pi/8\ln 2)^{1/4}}

This appendix considers the detectability of pulsars with respect
to two broad search techniques: one that exploits the inherent periodicity
and another that identifies individual, large-amplitude pulses.
Burns \& Clark (1969) made a brief comparison of these techniques but, 
for the most part, pulsar astronomers have relied on periodicity searches.
Consider a time interval of length $T$ in which there are $N_p$
pulse periods.   Pulse intensity maxima
are described by a continuous probability density function (PDF)
$\pdfi(S)$ and its integral, the cumulative distribution function (CDF)
$\cdfi(S)$.

\subsection{ Single-Pulse Search Detection}

Given system noise $\ssys$ expressed
in Janskys, the signal-to-noise ratio (S/N) of the pulse peak $\im$ in the
sum of $\npol$ polarization channels when the
pulse is match filtered is
\be
\left ( \snr \right )_{\rm SP} = \eta (\npol\Delta\nu W)^{1/2} \ssys^{-1} \im,
\ee
where $\eta\sim 1$ is a pulse-shape-dependent factor, $\Delta\nu$ is the total bandwidth and
$W$ is the pulse width.
For a Gaussian pulse shape,
$\eta = \snrpkfactor \approx 0.868$.
Single pulse detection would entail specifying some threshold $\snr$
of at least 3 and probably much larger, depending on how many
statistical trials were made in a survey.

For PDFs with long tails, one should take into account
the fact that $\pdfi$ is truncated at the maximum likely intensity in
the interval.
Assuming that all pulses have the same
width $W$, we define the maximum pulse intensity likely to occur in
the interval from the constraint $[1-\cdfi(\im)] N_p = 1$, implying
\be
\cdfi(\im) = 1 - N_p^{-1}.
\ee

\subsection{Periodicity Search Detection}

In a search for a periodic signal in the time series, the Fourier analysis
effectively measures the period-averaged mean intensity for the time interval.
The period-averaged intensity is defined to be
$\iavephi = \zeta (W/P)\iave$, where $\iave$ is the mean {\em peak} intensity,
$\zeta$ is a shape-dependent constant $\sim 1$ and $P$ is the pulse period.  For a
Gaussian pulse with FWHM $W$, $\zeta = \sqrt{\pi / \ln 2}/2 \approx 1.06$.
In terms of the pulse-peak PDF,
the ensemble mean of the peak intensity is
\be
\iave \equiv \int dS\, S \pdfi(S).
\ee
However, recognizing that PDFs with long tails yield a maximum intensity
in a finite number of pulse periods,
we define a modified mean intensity,
\be
\iavep 	= \frac
	 {\int_{0}^{\im}dS\, S \pdfi(S)}
	 {\int_{0}^{\im}dS\,  \pdfi(S)}
	= \left ( 1 - N_p^{-1}\right)^{-1}
		\int_{0}^{\im}dS\, S \pdfi(S).
\ee

The \snr\ for a periodicity search over an interval
$T \equiv N_p P$ is
\be
\left ( \snr \right )_{\rm FFT} =
	(\npol\Delta\nu T)^{1/2} \ssys^{-1}  \iavephi\ \hsum,
\label{eq:psnr}
\ee
where the harmonic sum is
\be
\hsum = N_h^{-1/2} \sum_{\ell = 1}^{N_h} R_{\ell},
\ee
with $R_{\ell}$ equal to the ratio of the $\ell$-th harmonic's
Fourier amplitude to that of the DC value.
The leading factor $N_h^{-1/2}$ takes into account that the
threshold increases as the number of harmonics summed increases.
For a Gaussian pulse shape,
\be
R_{\ell} = \exp \left [
		-\left ( \frac{\pi W \ell}{2 P \sqrt{\ln 2}} \right)^2
		\right],
\label{eq:gausspulse}
\ee
and the number of harmonics maximizing $\hsum$ is
${N_h}_{\rm max} \approx P / 2W$.
The value of the harmonic sum at the maximum is well approximated
by
${\hsum}_{\rm max} \approx 1/2 \left ( P/W \right)^{1/2}$.
The $\snr$ of the periodicity search using the harmonic sum is then
\be
\left ( \snr \right )_{\rm FFT} &=&
	(\zeta / 2)
	(\npol\Delta\nu W)^{1/2} \ssys^{-1}  N_P^{1/2} \iavep.
\ee

\subsection{Comparing Single-Pulse and Periodicity Detection}

A comparison of the two techniques is best done by considering the ratio
of the two signal-to-noise ratios:
\be
r &\equiv& \frac
		{\left ( \snr \right )_{\rm SP}}
		{\left ( \snr \right )_{\rm FFT}}
 \nonumber \\
  &=&
	\left (
                \frac{2\eta }{\zeta N_p^{1/2}}
        \right )
	\frac
	 {\im}
	 {\iavep}
\label{eq:ratio2}
\nonumber \\
  &=&
	\left (
                \frac{1}{N_p}
        \right )^{1/2}
	\frac
	 {2\eta\im}
	 {\zeta\left ( 1 - N_p^{-1}\right)^{-1}
		\int_{0}^{\im}dS\, S \pdfi(S)}.
\label{eq:ratio3}
\ee

\subsection{Unimodal Distributions}

Inspection of Eq.~\ref{eq:ratio2} indicates that in order for
$r>1$, $\im/\iavep > \zeta N_p^{1/2}/2\eta$, which is a large number
for typical observation times $T$ and pulse periods $P$.
Some unimodal PDFs satisfy this criterion for large $N_p$ while many do not.

We first
consider an {\it exponential} PDF,
\be
\pdfi(S) = \iave^{-1} e^{-S/\iave}U(S),
\ee
where $U$ is the unit step function.  For this case
the maximum pulse intensity encountered in $N_p$ trials is
\be
\im = \iave \ln N_p,
\ee
the renormalized average is
\be
\iavep = \iave \left [
	1 - \frac{\ln N_p}{N_p - 1}
		\right ]
\ee
and ratio of  $\snr$ for the two search methods becomes
\be
r =
	\frac
		{2\eta\ln N_p}
		{\zeta N_p^{1/2} \left[1 - \ln N_P/(N_p-1)\right]}.
\ee
For this case, $r>1$ only for $N_p \le 47$.

We next consider a {\it power law} PDF.   A truncated power law PDF can be
written in the form
\be
\pdfi(S) &=& {\cal N} S^{-\alpha}, \quad S_1 \le S \le S_2 \\
{\cal N} &=& \left\{
		\begin{array}{ll}
			(\ln S_2/S_1)^{-1} &
				\quad\quad \mbox{$\alpha = 1$} \\
			\\
			\displaystyle
         		\frac{(1-\alpha)} {(S_2^{1-\alpha} - S_1^{1-\alpha})} &
				\quad\quad \mbox{$\alpha \ne 1$}.
		\end{array}
		\right.
\ee
For a flat PDF ($\alpha = 0$) all pulse intensities are equally probable,
so we expect that as $N_p$ gets large, the maximum and minimum intensities
will equal the cutoff values. Letting $\im \to S_2$, we find
\be
r =
        \frac{4\eta}{\zeta N_p^{1/2}}
        \frac {1}{\left (1 + S_1 / S_2\right)}.
\ee
Asymptotically,  the maximum $r$ results for $S_2/S_1 \gg 1$,
\be
r_{\rm max} \to
         \frac{4\eta}{\zeta N_p^{1/2}},
\ee
which yields $r > 1$ only for $N_p < 11$.

We will not consider power laws
with $\alpha < 0$.  For $\alpha > 0$,  larger amplitudes are less
probable, so we expect that the maximum encountered out of $N_p$ pulses
generally will be less than $S_2$.   We find that
\begin{equation}
\im = \left \{
	\begin{array}{ll}
	  S_1 \left(\displaystyle\frac{S_2}{S_1}\right)^{1-N_p^{-1}}, & \quad \mbox{$\alpha = 1$} \\
	  \\
	  \displaystyle\frac{S_2}
	  {\displaystyle
	    \left[
	    \left(\frac{N_p-1}{N_p}\right)
	    \left(
		1 + \left( \frac{S_2}{S_1} \right)^{\alpha-1} \frac{1}{N_p-1}
	    \right)
	    \right]^{\displaystyle\frac{1}{\alpha-1}}
	  }
	  & \quad \mbox{$\alpha \ne 1$}
	\end{array}
	\right.
\label{eq:im}
\end{equation}
Using $\im$ we evaluate\footnote{Note that we evaluate
$\iavep$ by integrating from the actual lower cutoff $S_1$ to
the effective upper cutoff $\im$.  This is valid only for monotonically
decreasing PDFs, i.e. $\alpha >0$.  For a flat PDF, moreover, we should
let $\im \to S_2$.}
$\iavep$
\begin{equation}
\iavep = \left \{
        \begin{array}{ll}
	\displaystyle
	  \frac{\im-S_1}{\ln \im / S_1}
		& \quad \mbox{$\alpha = 1$} \\
          \\
	\displaystyle
	  \frac{S_1 \im \ln \im / S_1} {\im - S_1}
		& \quad \mbox{$\alpha = 2$} \\
          \\
	\displaystyle
          {\im} \left( \frac{\alpha-1}{\alpha-2}\right)
            \left[
	    \frac
		{ \left(\displaystyle\frac{\im}{S_1}\right)^{\alpha-2}-1}
		{ \left(\displaystyle\frac{\im}{S_1}\right)^{\alpha-1}-1}
            \right]
          	& \quad \mbox{$\alpha \ne 1,2.$}
        \end{array}
        \right.
\label{eq:iavep}
\end{equation}

For steeper power laws with $\alpha > 0$, the
\snr\ ratio $r>1$ only for small $N_p$ for $\alpha \lesssim 1$.
For intermediate cases, $1.5 \lesssim \alpha \lesssim 2.5$,
$r$ has a maximum at $N_p \gg 1$.   For steep power laws
(e.g. $\alpha \ge 3$) the ratio $r>1$ again only for small
$N_p$ and decreases monotonically.

Figure~\ref{fig:powerlaws} shows $r$ for several power-law cases and for
the exponential PDF and illustrates the statements made in the
preceding paragraph.   The values for $r$ are small for nearly flat PDFs
because the pulses Fourier-analyzed in a long-train of pulses have
amplitudes that do not deviate much from the peak pulse.   For very
steep power laws (e.g. $\alpha \ge 3$),  the probability of
seeing a large pulse near the upper cutoff, $S_2$, is too small to outweigh
the $N_p^{1/2}$ increase in \snr\ of the Fourier method, even for very
large $N_p$.   For intermediate cases, where $r$ has a distinct maximum,
the likelihood of seeing  a large pulse outweighs
the $N_p^{1/2}$ increase in \snr\ of the Fourier method.

\medskip
\epsfxsize=9truecm
\epsfbox{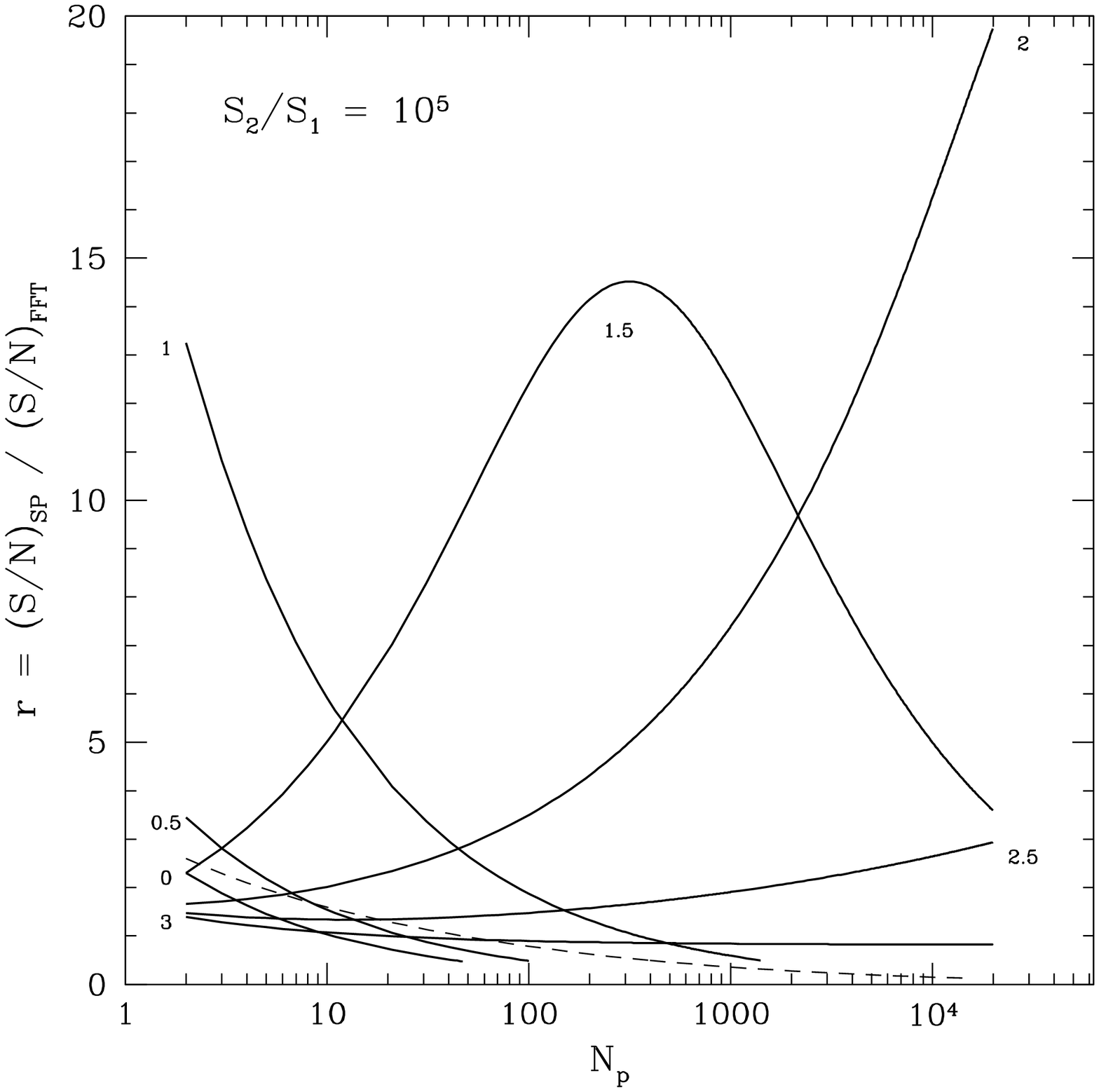}
\figcaption{
The \snr\ ratio $r$ for power-law intensity PDFs (solid lines) and
for an exponential PDF (dashed line).    The exponent of the PDF $\alpha$
is shown and all power-laws are for a ratio of cutoff intensities
$S_2/S_1 = 10^5$.  Single-pulse detection is superior to a Fourier harmonic-sum
detection scheme if $r > 1$.
\label{fig:powerlaws}
}
\bigskip

For $\alpha > 1$ and $\alpha \ne 2$ we  can express $r$ in the
limit where $S_2 / S_1 \gg 1$ using Eq.~\ref{eq:im}-\ref{eq:iavep}:
\be
r \approx \left(\frac{2\eta}{\zeta}\right)
	  \left( \frac{\alpha-2}{\alpha-1} \right) \times
	  \left\{
		\begin{array}{ll}
		\displaystyle
		   N_p^{\displaystyle\frac{3-\alpha}{2(\alpha-1)}}, &
			\quad\quad \mbox{$N_p \ll (S_2/S_1)^{\alpha-1}$} \\
		   \\
		   \left(\displaystyle\frac{S_2}{S_1}\right) N_p^{-1/2}, &
			\quad\quad \mbox{$N_p \gg (S_2/S_1)^{\alpha-1}$}.
		\end{array}
	\right.
\label{eq:scalings}
\ee
For small $N_p$, $r$ scales as
$r\propto N_p^{(3-\alpha)/2(\alpha-1)}$
until $\im \to S_2$ and then $r$ decreases as $N_p^{-1/2}$.
Note that $r$ has an increasing branch only if $1 < \alpha < 3$,
consistent with Figure~\ref{fig:powerlaws}.   If there is an increasing
branch, then we expect a maximum  approximately where the two
scaling laws in Eq.~\ref{eq:scalings} are equal:
\be
{N_p}_{\rm max} \approx \left(\frac{S_2}{S_1}\right)^{\alpha-1}.
\ee
For $S_2/S_1 = 10^5$, we expect the maximum for
$\alpha = 3/2$ to be at $N_p \approx 10^{5/2}$, consistent with
the exact expressions evaluated in the Figure.  The maxima for
$\alpha = 2$ and $\alpha=5/2$ are then expected at
${N_p}_{\rm max} \approx 10^5$ and $10^{15/2}$ and are thus off the scale
of Figure~\ref{fig:powerlaws}.

{\it We conclude that for certain power laws, intensity cutoffs and numbers of pulses,
single-pulse searches can be superior to periodicity searches.}

For the Crab pulsar at 146 MHz, Argyle \& Gower (1972) find
$\alpha \approx 2.5$ over at least two orders of magnitude of
flux density (c.f. Figure 4-9 of Manchester \& Taylor 1977).
This slope is consistent with the fact that the Crab pulsar is more
easily detected (and was discovered through) using its giant pulses.
At 812 MHz Lundgren et al. (1995) find $\alpha \approx 3.5$ over
1.2 orders of magnitude.   They also infer that the PDF for all pulses
(giant and normal) is bimodal.   Therefore, even though the slope found
by Lundgren et al. would suggest that single-pulse detection would be
inferior to a periodicity search, the apparent bimodality appears to change
this conclusion, as we show below.

\subsection{Bimodal Distribution}  

A bimodal distribution provides a finite
phase space for $r>1$, even for large pulse numbers $N_p \gg 1$.
Consider the bimodal distribution
\be
\pdfi(S) = (1-g)\delta(S - S_1) + g\delta(S - S_2).
\ee
The only interesting case is when the interval contains at least one
pulse with intensity $S_2$, requiring $gN_p \gtrsim 1$ or
$g > g_{\rm min}$ where $g_{\rm min} = N_p^{-1}$.     In this
case we have
\be
r =
	\frac{2\eta S_2}
	{\zeta N_p^{1/2} \left [ (1-g)S_1 + g S_2 \right ]}.
\ee
To have $r>1$ requires $g_{\rm min} < g < g_{\rm max}$ where
\be
g_{\rm max}  =  \frac
	{
		(2\eta/\zeta)N_p^{-1/2}  - S_1 / S_2
	}
	{1 - S_1  / S_2}.
\ee
Figure \ref{fig:regimes}
shows regions in the two dimensional space of $g$ and $S_1/S_2$
for which $r>1$.   The regions are a function of $N_p$.

\medskip
\epsfxsize=9truecm
\epsfbox{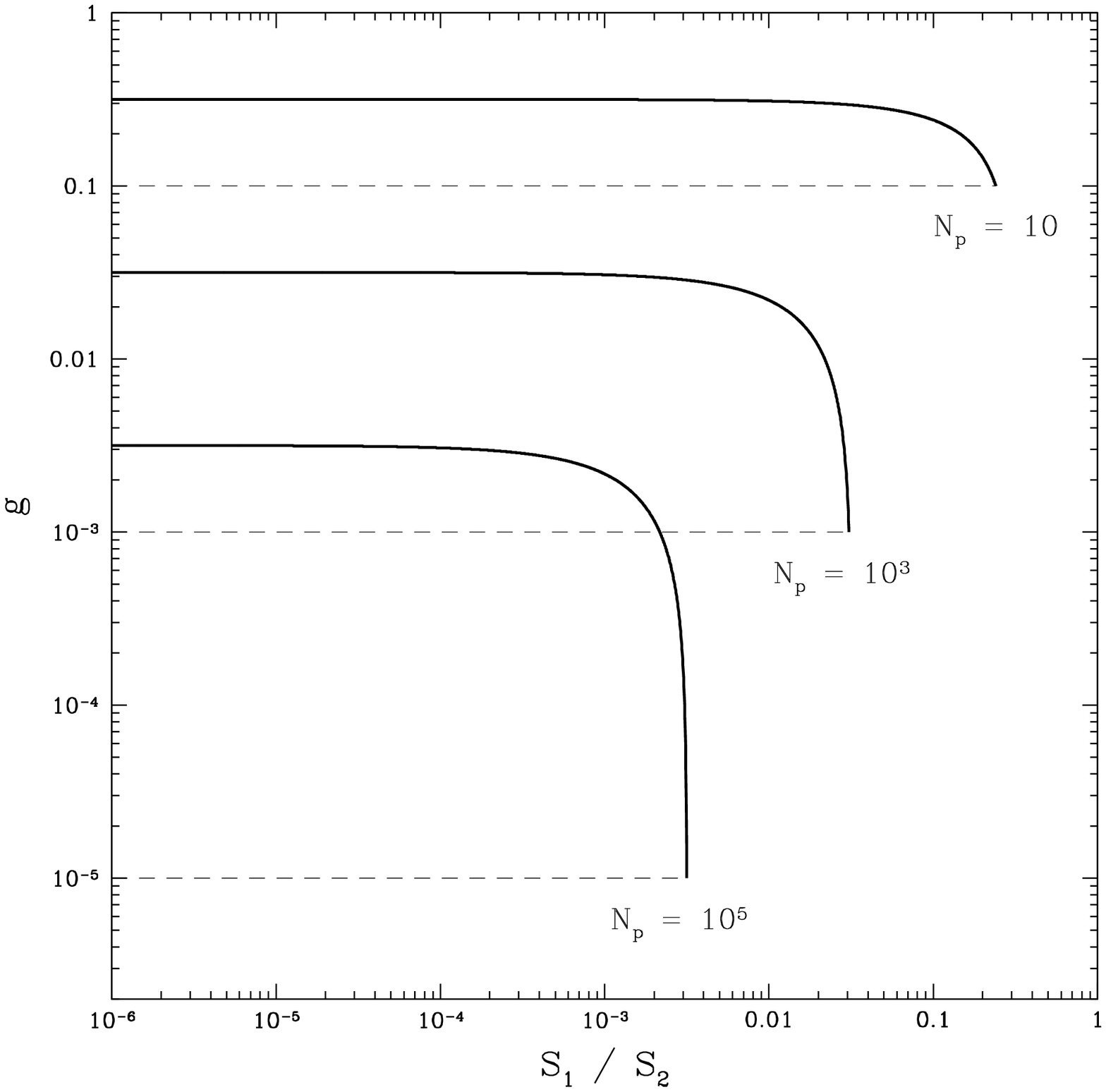}
\figcaption{
Domain for which a single-pulse search is more sensitive than
a periodicity search.  The pulse intensity distribution is assumed
to be bimodal.
$S_1$ is the intensity of ordinary pulses with probability $1-g$
and $S_2$ is the giant-pulse intensity with probability $g$.
The results depend on how many pulse periods, $N_p$, are analyzed.
For each of the three cases
shown, a single-pulse search is superior for values  of $g$ and
$S_1/S_2$ between the dashed and solid lines.
\label{fig:regimes}
}
\bigskip

A special case of interest  for the bimodal PDF
is where the lowest intensity pulses are null pulses
(i.e. $S_1 = 0$).   Then
\be
r =
        \frac{2\eta}
        {g\zeta N_p^{1/2}}.
\ee
Single pulse detection is superior to a periodicity search
when
\be
N_p^{-1} < g  < (2\eta/\zeta) N_p^{-1/2}.
\ee
This regime corresponds to values of $g$ between the solid and dashed
lines in Figure \ref{fig:regimes} for small values of $S_1/S_2$.

Under some other special circumstances, searches for single pulses are more effective
than periodicity searches.  
For large duty cycles,
the single-pulse detection scheme becomes exponentially
better once the pulse width exceeds the pulse period.    Using the
general expressions in Eq.~\ref{eq:psnr} through
Eq.~\ref{eq:gausspulse} when
calculating $r$ as a function of duty cycle, $W/P$, we find as a typical
trend that the periodicity search is superior for $W/P \lesssim 1$
and if the single-pulse amplitude PDF does not have a large dynamic range
in amplitudes, as described in the previous subsection.     However,
for $W/P \gtrsim 1$, the single-pulse search is generally superior,
no matter the form for the amplitude PDF.

These results imply that sources with small spin and small orbital periods
may be detected in single-pulse searches when they would be missed
completely in periodicity searches.   Similarly, distant pulsars viewed
through deep scattering volumes and thus having large
scattering measures are likely to be missed in periodicity searches.
In this case, the scattering time is a strong function of frequency, and so too will
be the relative sensitivities of single-pulse and periodicity searches.

\clearpage

\begin{deluxetable}{lccc}
\tablewidth{4.5in}
\tablenum{1}
\tablecaption{Radio Pulsars with Highest $B_{lc}$}
\tablehead{
\colhead{Name} & \colhead{$P$ (ms)} & \colhead{$\dot{P}$ (10$^{-15}$ s s$^{-1}$)} & \colhead{$B_{lc}$ ($10^5$ Gauss)} }
\startdata
B1937+21 & 1.56 & $1.1 \times 10^{-4}$ & 9.8 \\
B0531+21 & 33.4 & $4.2 \times 10^{2}$ & 9.3 \\
B1821$-$24 & 3.05 & $1.6 \times 10^{-3}$ & 7.2\\
B1957+20 & 1.61 & $1.7 \times 10^{-5}$ & 3.6 \\
B0540$-$69 & 50.4 & $4.8 \times 10^{2}$ & 3.5 \\
J0218+4232 & 2.32 & $7.5 \times 10^{-5}$ & 3.1 \\
J1823$-$3021A & 5.44 & $3.4 \times 10^{-3}$ & 2.5 \\
J0034$+$0534 & 1.88 & $6.7 \times 10^{-6}$ & 1.6 \\
J2229+6114 & 51.6 & $7.8 \times 10^{1}$ & 1.3 \\
J0205+6449 & 65.7 & $1.9 \times 10^{2}$ & 1.2 \\
\enddata
\label{tab:blcs}
\end{deluxetable}{}

\clearpage

\begin{deluxetable}{lcccc}
\tablewidth{3.5in}
\tablenum{2}
\tablecaption{Pointings at M33}
\tablehead{
\colhead{$N$} & \colhead{RA} & \colhead{DEC} & \colhead{$N_p$} & \colhead{Total time (s)}}
\startdata
1 & 01:33:51 & 30:59:52 & 7  & 8100 \\
2 & 01:33:37 & 30:53:07 & 6  & 7800 \\
3 & 01:34:04 & 30:53:07 & 8  & 10800 \\
4 & 01:33:24 & 30:46:22 & 8  & 10800 \\
5 & 01:33:51 & 30:46:22 & 8 & 10800  \\
6 & 01:34:18 & 30:46:22 & 5  & 6300 \\
7 & 01:33:10 & 30:39:37 & 7  & 9000 \\
8 & 01:33:37 & 30:39:37 & 6 &  6300 \\
9 & 01:34:04 & 30:39:37 & 8  & 8100 \\
10 & 01:34:31 & 30:39:37 & 10 & 9900  \\
11 & 01:33:24 & 30:32:52 & 8  & 8100 \\
12 & 01:33:51 & 30:32:52 & 6  & 7200 \\         
13 & 01:34:18 & 30:32:52 & 7  & 7800 \\
14 & 01:33:38 & 30:26:07 & 7 & 8100  \\
15 & 01:34:04 & 30:26:07 & 6 & 7200 \\
16 & 01:33:51 & 30:19:22 & 7  & 8100 \\
off1 & 00:33:00 & 30:59:52 & 2  & 3600 \\
off2 & 02:33:51 & 30:59:52 & 3  & 3600 \\
\enddata
\label{tab:m33beams}
\end{deluxetable}{}

\clearpage

\begin{deluxetable}{lcccccc}
\tablewidth{5.0in}
\tablenum{3}
\tablecaption{Pointings at Other Galaxies}
\tablehead{
\colhead{Name} & \colhead{RA} & \colhead{DEC} & \colhead{$l$} & \colhead{$b$} & \colhead{$D$ (Mpc)} & \colhead{N$_{p}$}}
\startdata
NGC253 & 00:47:33.1 & $-$25:17:18 & 97.3 & $-$88.0 & 3 & 3 \\
NGC300 & 00:54:53.5 & $-$37:41:00 &  299.2 & $-$79.4 & 2 &  2 \\
Fornax  & 02:39:34 & $-$34:31:08 & 237.3 & $-$65.7 & 16.9 & 1 \\
LMC & 05:33:14\tablenotemark{\dagger} & $-$69:21:38$^{\dagger}$ &  279.4 & $-$32.17 & 0.050 &  6 \\
NGC6300 & 17:16:59.2 & $-$62:49:11 &  328.5 & $-$14.1 & 16.9 & 1 \\
NGC7793 & 23:57:49 & $-$32:35:24 &  4.6 & $-$77.2 & 3.5 &  3 \\ \hline
\enddata
\tablenotetext{\dagger}{Coordinates of the central LMC pointing.}
\label{tab:othergal}
\end{deluxetable}

\end{document}